\documentclass[11pt, reqno]{amsart}
\usepackage{amsmath}
\usepackage[margin=1.00in]{geometry}
\usepackage[foot]{amsaddr}
\usepackage{amssymb}
\usepackage{mathrsfs}
\usepackage{enumerate}         %
\usepackage{color}
\usepackage{cases}
\usepackage{url} 
\usepackage{amsthm}
\usepackage{hyperref}
\hypersetup{hidelinks}
\usepackage{bm}
\usepackage{xy}
\usepackage{enumitem}
\usepackage[gen]{eurosym}
\usepackage{graphicx}
\usepackage{tikz}
\usetikzlibrary{positioning,arrows.meta}

\usepackage{bm}
\usepackage{booktabs}
\usepackage{makecell}

\usepackage{orcidlink}

\usepackage{natbib}

\theoremstyle{plain}

\theoremstyle{definition}

\theoremstyle{remark}
\newtheorem{remark}{Remark}[section]

\title[Regime Detection in Latent Factor Models]{Regime Detection in Latent Factor Models\\  -- A Bayesian Decision-Theoretic Approach}

\author{Karl Kristian Engelund\orcidlink{0009-0000-2182-666X}\\\tiny Dept. of Mathematical Sciences, University of Copenhagen\\ Copenhagen, Denmark}
\email{kken@math.ku.dk}
\date{December 1, 2026}

\usepackage{etoolbox}

\makeatletter
\pretocmd{\@setabstract}{%
    \vspace{0.2cm}
    \begin{center}
        \small
        This article has been accepted for publication in 
        {\itshape Quantitative Finance}, published by Taylor \& Francis.
    \end{center}
}{}{}
\makeatother

\begin{document}

\begin{abstract}
Regime-switching multi-factor models are widely used in quantitative finance, but filtering does not by itself determine when an agent should act on inferred regime changes. We formulate online regime detection with latent factors and hidden regimes as a Bayesian decision problem balancing false alarms, detection delay, and misclassification. Since posterior regime probabilities are generally non-Markovian, the problem is represented as a measure-valued optimal stopping problem and approximated using particle filtering with Monte Carlo dynamic programming. A single-change-point Heston model provides a benchmark. With observed variance, the problem reduces to a two-dimensional Markov free-boundary problem, while under latent variance decisions are based on the filtered posterior measure. Numerical experiments in Heston and regime-switching Vasicek models show that the latent-information procedure can outperform feasible plug-in rules.
\end{abstract}

\maketitle

	\vspace{-0.5cm}

	

\section{Introduction}

Factor models are standard tools in quantitative finance for describing asset returns, volatility dynamics, and interest rates. In equity and foreign exchange option markets, stochastic volatility models such as the Heston model provide a reasonable fit to S\&P 500 option data \citep[see, e.g.,][]{heston1993closed,nandi1998important}. In fixed income and commodity markets, affine multi-factor term-structure models capture important features of swap and derivatives data, including evidence of unspanned volatility factors \citep[see, e.g.,][]{duffie1996yield,dai2000specification,trolle2009unspanned}. Similar latent-factor representations also arise in market microstructure, where variables such as liquidity and order flow evolve stochastically over time \citep{collin2015prices}. Beyond empirical fit, these models are attractive because they often admit tractable pricing formulas or efficient numerical approximations \citep[see, e.g.,][]{carr1999option,duffie2000transform}.

At the same time, the limitations of constant-parameter factor models are well documented. In equity derivatives, models in the Heston class can struggle to capture the term structure of implied volatility and to jointly calibrate SPX and VIX options \citep[see, e.g.,][]{pan2002jump,kokholm2015joint}. In other markets, abrupt changes in volatility or order-flow dynamics suggest structural shifts in the underlying state variables \citep[see, e.g.,][]{tsaknaki2025online,johansen2026arithmetic}. A natural extension is therefore to introduce hidden market regimes and allow the parameters of the factor model to switch with the regime.

Regime-switching models have a long history in statistics, econometrics, and finance, with early contributions including \cite{quandt1958estimation}, \cite{GOLDFELD19733}, and \cite{tong1983threshold}, followed by the seminal paper by \cite{hamilton1989new}. %
Subsequent studies document regime changes in asset dynamics and develop pricing methodologies under hidden regimes \citep[see, e.g.,][]{engel1990long,naik1993option,gray1996modeling,duan2002option,mikosch2004changes,elliott2005option,SIU2008295,Elliott2009,e69e7ccee7bf47feab824f343699e4bd}.  %

In option markets, regime-switching stochastic volatility models have been shown to improve joint calibration to SPX and VIX option data \citep{Papanicolaou01102014,goutte2017regime}. In fixed income, regime-switching term-structure models can reduce the number of factors required to match U.S. Treasury data \citep{christensen2013regime}. Because these models contain both latent factors and hidden regimes, much of the existing literature focuses on estimation and calibration. Standard approaches include Kalman filtering in linear Gaussian settings, extended Kalman filtering, and particle-filtering methods for more general nonlinear and non-Gaussian models \citep[see, e.g.,][]{babbs1999kalman,christensen2013regime,kang2014estimation,doucet2011tutorialPF,mcginnity2002multiple,carvalho2007simulation}. More recent work develops particle filters that are more robust to regime changes or amenable to machine-learning-based training \citep{gellert2021parameter,brady2026differentiable}.

\begin{figure}[h]
    \centering
    \begin{tikzpicture}[
        node distance=14mm,
        >=Latex,
        block/.style={draw, rounded corners, align=center, inner sep=6pt, minimum width=34mm},
        small/.style={draw, rounded corners, align=center, inner sep=5pt, minimum width=34mm},
        arrow/.style={->, thick},
        dashedarrow/.style={->, thick, dashed},
        doublearrow/.style={<->, thick},
        dotblock/.style={block, dotted},
        dotarrow/.style={->, dotted}
    ]
        \node[block] (Z) {Hidden regime\\ $Z_t$};

        \node[block, below right=2mm and 5mm of Z] (X) {Latent factors\\ $\boldsymbol{X}_t$};
        \node[block, above right=2mm and 5mm of Z] (Y) {Observed factors\\ $\boldsymbol{Y}_t$};

        \node[block, right=8mm of Y] (filter) {Filtering\\
        $\boldsymbol{X}_t, Z_t$ given $\mathcal{F}^{\boldsymbol{Y}}_t$};
        \node[block, right=8mm of filter] (dec) {Decision rules\\ $(\tau,d)$};
        \node[dotblock, below=8mm of dec] (act) {Action \\ Rebalance, hedge, etc.};

        \draw[arrow] (Z) -- (X.west);
        \draw[arrow] (Z) -- (Y.west);
        \draw[doublearrow] (X) -- (Y);

        \draw[arrow] (Y) -- (filter);
        \draw[arrow] (filter) -- (dec);
        \draw[dotarrow] (dec) -- (act);

        \draw[dashedarrow] (filter.south) to[out=-130,in=15]  (X.east);
        \draw[dashedarrow] (filter.south) to[out=-170,in=0]  (Z);
    \end{tikzpicture}
    \caption{Schematic of the decision-theoretic problem in regime-switching multi-factor models. Filtering draws inference on the latent factors and the hidden regime, which is then used to construct a decision rule. The downstream actions taken based on these decisions are outside the scope of this paper.}
    \label{fig:schematic}
\end{figure}
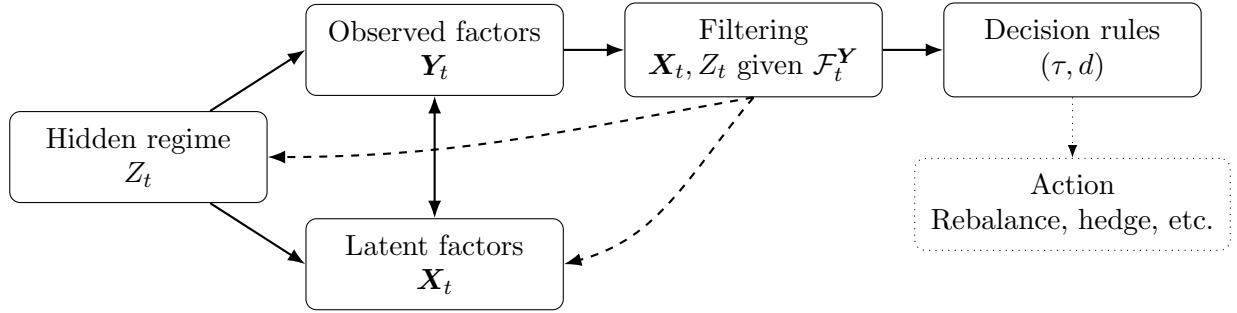

In contrast, explicit regime detection in a decision-theoretic sense has received much less attention in quantitative finance. Filtering methods provide sequential estimates of latent regimes and factors, but they do not by themselves specify when a financial agent should act on those estimates. In many applications, this distinction matters. Portfolio rebalancing, hedging adjustments, or changes in trading behavior may be costly, so acting continuously on filtered regime probabilities is generally not optimal. This motivates a Bayesian decision-theoretic formulation in which a decision consists of a stopping time and a regime declaration, chosen to minimize expected loss from false alarms, delayed detection, and misclassification, see the schematic in Figure \ref{fig:schematic}. In the classical setting, sequential testing and quickest detection problems are well understood because a suitable posterior probability process is Markov and the problem reduces to a low-dimensional optimal stopping problem \citep[see, e.g.,][]{shiryaev2008optimal,gapeev2013bayesian,johnson2017quickest,glover2024quickest,ernst2024gapeev}. For more general regime-switching models, the corresponding posterior regime probabilities are typically not Markov, and exact Bayes rules are no longer available in tractable form \citep[see, e.g.,][]{ludkovski2012bayesian}.

This paper studies online regime detection in regime-switching multi-factor models under partial observation. It adopts a Bayesian decision-theoretic framework and formulates %
the problem as an optimal stopping problem driven by filtered information. Our starting point is a single-change-point Heston model, which serves as a simple but financially relevant benchmark. Under an idealized oracle formulation in which the variance process is observed, the problem becomes finite-dimensional and can be characterized by standard optimal stopping methods. This provides a useful benchmark for the latent-information setting.

When the variance process is treated as latent, we formulate the state by the posterior measure on the latent factor and regime, leading to a measure-valued optimal stopping problem. This measure-valued state is infinite-dimensional in general regime switching multi-factor models, although it may admit a finite-dimensional representation in special cases such as the ideal continuous-observation Heston model.

To obtain implementable decision rules for measure-valued optimal stopping problems, we resort to a particle-filter Monte Carlo dynamic programming (PFMCDP) %
based approach. \citet{ludkovski2009simulation} formulates partially observed optimal stopping using the conditional distribution as a Markov state and combines particle filtering with regression Monte Carlo, including an error analysis. This has, for example, been applied to American option pricing under latent stochastic volatility. Moreover, \citet{ludkovski2012bayesian} applies related filtering and Monte Carlo dynamic programming ideas to Bayesian quickest detection in sensor arrays. Thus, we combine the two approaches for our decision-theoretic regime detection problem in regime-switching multi-factor models, where continuous latent factors and a hidden Markov regime are filtered jointly.

Two numerical experiments are considered. The first considers a regime-switching Vasicek model with measurement-noise-contaminated bond prices and is used to study sensitivity to the main approximation parameters. The second revisits the single-change-point Heston model and compares the latent variance procedure with the oracle rule and several feasible plug-in benchmarks, including realized-variance- and filter-based proxy rules. 
At the sampling frequencies considered, the PFMCDP method yields lower estimated loss than the feasible proxy rules, with a statistically significant improvement over the particle-filter plug-in rule.

The contribution of the paper is therefore twofold. First, it formulates online regime detection in regime-switching multi-factor models as a Bayesian decision-theoretic problem relevant for financial decision making.
Building on existing particle-filter and regression Monte Carlo methods for partially observed optimal stopping, it applies this methodology to the aforementioned model and evaluates the resulting decision rules against oracle and plug-in benchmarks in the single-change-point Heston example. Specifically, using the conditional-likelihood construction lets the PFMCDP and the particle-filter plug-in rule use the same posterior estimate, isolating the effect of optimizing the stopping decision under partial information.
Second, to the author's knowledge, this single-change-point Heston regime-detection problem has not been studied previously. In the oracle formulation, joint observation of the correlated stock and variance processes leads to the two-dimensional Markov state $(\nu_t,\pi_t)$, whose diffusion is locally elliptic for $\rho\neq0$ and degenerate but hypoelliptic for $\rho=0$. We derive the corresponding free-boundary characterization and analyze the problem.

The remainder of the paper is organized as follows. Section \ref{sec:heston} introduces the single-change-point Heston model and studies both the oracle and latent variance formulations. Section \ref{sec:general_setup} develops the general regime-switching multi-factor framework and the associated measure-valued optimal stopping problem. Section \ref{sec:num_theory} presents the numerical approximation procedure, and Section \ref{sec:num} reports the numerical experiments. Additional details on the oracle problem, filtering equations, particle filter, and Vasicek bond-pricing model are collected in the appendices.

\section{Regime detection in a single-change-point Heston model}\label{sec:heston}%
We begin with a simple model to build intuition for the features of the full regime-switching multi-factor model. Let $(\Omega,\mathcal{F},(\mathcal{F}_t)_{t\geq0},\mathbb{P})$ be a complete %
filtered probability space supporting a two-dimensional Brownian motion $(W_t,B_t)$ with instantaneous correlation $\mathrm{d}\langle W,B\rangle_t=\rho\,{\rm d}t$, for $\rho\in(-1,1)$. The traded asset $(S_t)$ follows a stochastic-volatility model under the physical measure $\mathbb{P}$, 
\begin{equation*}
    \mathrm{d}S_t=\mu S_t {\,\rm d}t+\sqrt{\nu_t}S_t{\,\rm d}W_t,
\end{equation*}
where $\mu\in\mathbb{R}$ and $(\nu_t)$ is the latent variance process with $\nu_t>0$. The filtered probability space also supports a continuous-time Markov chain $(Z_t)$, independent of $(W_t,B_t)$, with state space $\mathcal{S}=\{1,2\}$. We assume State $1$ is \emph{transient} and State $2$ is \emph{absorbing}. The initial distribution has prior $\mathbb{P}(Z_0=2)=1-\mathbb{P}(Z_0=1)=\pi$, and the hazard rate of transitioning from State 1 to State 2 is $\lambda\geq0$. %
When $\lambda = 0$, no future change-point occurs, but the initial regime is uncertain.
We consider a Heston model in which the change-point affects the mean-reversion level of the variance process such that
\begin{equation*}
    {\rm d}\nu_t=\kappa(m_{Z_t}-\nu_t){\rm d}t+\xi\sqrt{\nu_t}{\rm d}B_t,
\end{equation*} %
for constants $\kappa,m_i,\xi>0$ with $m_1\neq m_2$ satisfying the Feller conditions $2\kappa m_i>\xi^2$ for $i\in\mathcal{S}$ which ensure that $\nu_t>0$ for all $t\geq0$. %

It should be noted that the two-regime, single-change-point specification is adopted primarily for parsimony and to keep the idealized continuous-observation problem studied in Section \ref{sec:heston_oracle} illustrative. Economically, the transient State $1$ represents the pre-change regime, while the absorbing State $2$ represents a persistent structural change. For instance, taking $m_1<m_2$, State $1$ may represent a low long-run volatility regime. Once the change-point occurs, the model assumes a structural shift to a higher long-run volatility regime in State $2$ in which it remains over the relevant decision horizon. 
The restriction of only letting the change-point affect the mean-reversion level $m_i$ and not the speed of mean reversion $\kappa$ is again made for parsimony rather than necessity.
We also keep $\xi$ regime-independent because regime-dependent diffusion coefficients can make the regime directly identifiable under continuous observations by a similar argument to the one below. Nevertheless, the general model of Section \ref{sec:general_setup} removes these restrictions, allowing more general multi-factor change-point formulations. %

Only the path of the traded asset $(S_t)$ is observed. The Markov chain $(Z_t)$ governs the hidden volatility regimes expressed through different mean-reversion levels of the latent variance process $(\nu_t)$ affecting the diffusion of the asset $(S_t)$. The hidden regime is important for interpreting the current market state: is volatility temporarily elevated due to stochastic fluctuations, or has the system moved into a regime with a higher long-run variance level? More importantly, should this trigger an action, such as costly rebalancing? Before formulating the decision-theoretic objective, we discuss the identifiability of $(\nu_t)$.

Taking the quadratic variation of the log price gives us the integrated variance process which is adapted to the natural filtration of $(S_t)$. Differentiating gives
\begin{equation}\label{eq:qvar_diff}
    \frac{{\rm d}\langle \log S\rangle_t}{{\rm d}t} = \nu_t,
\end{equation}
which is adapted to the augmented filtration generated by $(S_t)$, denoted $(\mathcal{F}^S_t)$. This motivates considering $(S_t,\nu_t)$ as the observed pair, see Section \ref{sec:heston_oracle}.  %
In this continuous-observation setting, $\mathcal{F}^{S,\nu}_t=\mathcal{F}^{S}_t$. In applications, observations are available only at discrete time points. Obtaining \eqref{eq:qvar_diff} relies on the quadratic variation convergence, say
\begin{equation}\label{eq:qvar}
    \underset{n\to\infty}{\mathrm{plim}}\sum_{i=1}^{n} (\log S_{i\Delta }-\log S_{(i-1)\Delta }) ^2 = \int_0^t\nu_s {\rm d}s,
\end{equation}
for $\Delta=t/n$, e.g., through the local $M$-point realized variance estimator
\begin{equation}\label{eq:rvar}
    \mathrm{RV}^{(M)}_t = \frac{1}{M\Delta }\sum_{i=n-M+1}^n (\log S_{i\Delta }-\log S_{(i-1)\Delta }) ^2 \approx\nu_t.
\end{equation}
Under standard high-frequency conditions with $M\to\infty$ and $M\Delta\to0$, the local estimator \eqref{eq:rvar} attains the pointwise rate $O_p(\Delta^{1/4})$ when $M\asymp\Delta^{-1/2}$ \citep{jacod2013quarticity}. At practically available sampling frequencies, this convergence can still be slow, which motivates considering $(\nu_t)$ as a latent factor, see Section \ref{sec:heston_filter}.
Furthermore, in more general regime-switching models, the latent factors are not directly identifiable, so the latent-factor treatment of $(\nu_t)$ developed here helps the methodology of Section \ref{sec:general_setup}.

\subsection{Decision-theoretic objectives}\label{sec:heston_decision}
We formulate our objectives as classical Bayesian decision-theoretic objectives. Let $(Y_t)$ be the observed process (e.g., either $Y_t=(S_t,\nu_t)$ or $Y_t=S_t$). A decision $(\tau,d)$ consists of a $(\mathcal{F}^Y_t)$-stopping time $\tau$ and a $\mathcal{F}^Y_\tau$-measurable declaration $d$ with values in the state space of $(Z_t)$, i.e. $\mathcal{S}=\{1,2\}$. %
We restrict optimization to be over admissible stopping times for which the corresponding expected loss is finite throughout. %

\subsubsection*{Sequential testing.} %
If $\lambda=0$, then both states are absorbing and no change-point occurs. The problem then becomes a classical Bayesian sequential hypothesis testing problem. A canonical risk to minimize is a combination of a misclassification penalty and a time penalty, e.g.,
\begin{equation}\label{eq:st_general_risk}
    R^{\text{st}}(\tau,d)=a\mathbb{P}(Z_\tau\neq d)+c\mathbb{E}[\tau]
\end{equation}
where $a,c>0$ are constants. Since no change-point occurs, $\mathbb{P}(Z_\tau\neq d)=\mathbb{P}(Z_0\neq d)$. 

\subsubsection*{Quickest detection.} %
If $\lambda>0$, then we are in a classical quickest (single) change-point detection setting. Since only one change-point occurs, we will denote the transition time, including at 0, of $(Z_t)$ by $\theta\geq0$. We wish to decide when the change-point has occurred, that is the declaration is deterministically $d=2$ and we only consider the stopping time part of the decision.  The standard risk to minimize is a combination of the probability of false alarm and expected detection delay, e.g.,
\begin{equation}\label{eq:qd_simple_risk}
    R^{\text{qd}}(\tau)=b\mathbb{P}(\tau<\theta)+c\mathbb{E}[(\tau-\theta)^+]
\end{equation}
for some constants $b,c>0$, and where $(\cdot)^+=\max(\cdot,0)$.

These problems are examples of unconstrained decision problems. In some settings, one could also be interested in constraining the decisions. For instance, in the quickest detection problem one can consider minimizing the expected detection delay $\mathbb{E}[(\tau-\theta)^+]$ over admissible %
stopping rules $\tau$ such that the probability of false alarm does not exceed a threshold, 
\begin{equation*}
    \mathcal{T}_\alpha(\theta)=\{\tau\mid\mathbb{P}(\tau<\theta)\leq \alpha\}.%
\end{equation*}
It is easy to see that this constrained problem is closely related to \eqref{eq:qd_simple_risk} through the method of Lagrange multipliers  \cite[see, e.g.,][]{Shiryaev1963on}. 

Following \cite{robert2007bayesian}, we call the infimum of the risk over admissible decisions the Bayes risk, and an optimal decision realizing the infimum is called a Bayes rule. %
Let $(\pi_t)$ denote the posterior probability process, with $\pi_t=\mathbb{P}(Z_t=2\mid\mathcal{F}^Y_t)$ and $\pi_0=\pi$. For any fixed stopping time $\tau$, the optimal declaration $d^*$ in \eqref{eq:st_general_risk} is obtained by choosing the more likely state, namely $d^*=1$ if $1-\pi_\tau>\pi_\tau$ and $d^*=2$ otherwise. Hence,  
\begin{equation}\label{eq:V_st_gen}
    V^{\text{st}}=\inf_{(\tau,d)}R^{\text{st}}(\tau,d)=\inf_{\tau}\mathbb{E}\left[a\big((1-\pi_\tau)\wedge\pi_\tau\big)+\int_0^\tau c \,{\rm d}t\right],
\end{equation}
where $\wedge$ is the minimum operator, while \eqref{eq:qd_simple_risk} yields the Bayes risk
\begin{equation}\label{eq:V_qd_gen}
    V^{\text{qd}}=\inf_{\tau}R^{\text{qd}}(\tau)=\inf_{\tau}\mathbb{E}\left[b(1-\pi_\tau)+\int_0^\tau c\pi_t\,{\rm d}t\right].
\end{equation}
We will denote both of these problems by 
\begin{equation}\label{eq:OSP_general}
    V = \inf_\tau\mathbb{E}\left[g(\pi_\tau)+\int_0^\tau h(\pi_t){\rm d}t\right],
\end{equation}
with functions $g,h:[0,1]\to[0,\infty)$ corresponding to either \eqref{eq:V_st_gen} or \eqref{eq:V_qd_gen}. %

We have thus rewritten the decision problems as optimal stopping problems. However, solving \eqref{eq:OSP_general} still requires more structure, since the set of $(\mathcal{F}^Y_t)$-stopping times is very large and contains highly path-dependent rules. Some authors consider a smaller set of stopping times, namely first hitting times, i.e., threshold type strategies \cite[see e.g.][]{baras1987real}. But in order to ensure a Bayes rule is given by such a threshold type, we need to write the optimal stopping problem as a Markovian optimal stopping problem. Then one can invoke Bellman's optimality principle and get that the optimal stopping time only depends on the current state of the processes. This holds true even in classical problems when the signal is present in the drift of the observation process.

\subsection{Observed variance case}\label{sec:heston_oracle}
Assume that both the stock and variance paths are observed, i.e., $Y_t=(S_t,\nu_t)$, and let $\pi_t=\mathbb{P}(Z_t=2\mid\mathcal{F}^{S,\nu}_t)$. This is a sequential testing problem when $\lambda=0$ and a quickest detection problem when $\lambda>0$. These can be analyzed by modern optimal stopping techniques. %
We give the main results here and delegate details to Appendix \ref{app:oracle}.
We will denote the statewise drift by $\alpha_i(\nu)=\kappa(m_i-\nu)$. %
The posterior probability process $(\pi_t)$ is given by
\begin{equation}\label{eq:posterior_heston_joint}
    \pi_t=\frac{\pi+(1-\pi)\int_0^t\lambda e^{-\lambda s}L_s^{-1} {\rm d}s}{\pi+(1-\pi)\big( e^{-\lambda t}L_t^{-1}+\int_0^t\lambda e^{-\lambda s}L_s^{-1} {\rm d}s\big)},
\end{equation}
where $(L_t)$ is the likelihood process, 
see, e.g., \cite{cheridito2007market} for true martingality,  %
given by
\begin{equation}\label{eq:likelihood_heston_joint}
    L_t = \exp\left\{\int_0^t \eta_\rho(\nu_s){\rm d}B^{\infty,\perp}_s-\frac12\int_0^t\eta_\rho(\nu_s)^2{\rm d}s\right\}
\end{equation}
and $\eta_\rho$ is the correlation adjusted signal-to-noise ratio given by
\begin{equation*}
    \eta_\rho(\nu)=\frac{\kappa(m_2-m_1)}{\xi\sqrt{\nu}\sqrt{1-\rho^2}},
\end{equation*}
and $(B^{\infty,\perp}_t)$ is a Brownian motion under $\mathbb{P}^\infty$, independent of $(W_t)$, and can be calculated from $(Y_t)$, see \eqref{eq:bminf}.
By its definition as a posterior probability, $\pi_t\in[0,1]$ almost surely. %
In terms of dynamics, the posterior probability process is given by
$\pi_0=\pi$ and
\begin{equation}\label{eq:pi_innov}
    {\rm d}\pi_t=\lambda(1-\pi_t){\rm d}t+\eta_\rho(\nu_t)\pi_t(1-\pi_t){\rm d}I_t,
\end{equation}
where $(I_t)$ is the innovation Brownian motion under $\mathbb{P}$, independent of $(W_t)$, given by ${\rm d}I_t={\rm d}B_t^{\infty,\perp}-\eta_\rho(\nu_t)\pi_t\,{\rm d}t$.
Rewriting the dynamics of $(\nu_t)$ in terms of the innovation process, we get
\begin{equation}\label{eq:nu_innov}
    {\rm d}\nu_t=\bar\alpha(\nu_t,\pi_t){\rm d}t
    +\xi\sqrt{\nu_t}\big(\rho\,{\rm d}W_t+\sqrt{1-\rho^2}\,{\rm d}I_t\big),
\end{equation}
where 
\begin{equation*}
    \bar\alpha(\nu,\pi)=\kappa(m_1-\nu)+\pi\big(\kappa(m_2-m_1)\big).
\end{equation*}
With $(W_t,I_t)$ as the driving Brownian motion, equations \eqref{eq:pi_innov}--\eqref{eq:nu_innov} constitute a closed system for $(\nu_t,\pi_t)$, which is a strong Markov process in the augmented observation filtration $(\mathcal{F}^{S,\nu}_t)_{t\geq0}$. Its coefficients and the decision costs do not depend on the stock level, so $S_t$ is not an additional state variable. However, since the correlation adjusted signal-to-noise ratio is not constant, equation \eqref{eq:pi_innov} does not form a closed system for $(\pi_t)$ alone. This reveals that $(\pi_t)$ may not be a strong Markov process by itself. %

Thus, a threshold strategy based solely on $(\pi_t)$ cannot be expected to be a Bayes rule, while the methodology of Appendix \ref{app:oracle} can be used to describe a Bayes rule as a threshold strategy on $(\nu_t,\pi_t)$. Specifically, equation \eqref{eq:OSP_general} admits the Markovian formulation
\begin{equation}\label{eq:qd_heston_markov}
    V(\nu,\pi) = \inf_{\tau}\mathbb{E}\left[g(\pi_\tau)+\int_0^\tau h(\pi_u) {\rm d}u\;\Big\vert\; \nu_0=\nu, \,\pi_0=\pi\right].
\end{equation}

\begin{figure}[h]
    \centering
    \includegraphics[width=1\linewidth]{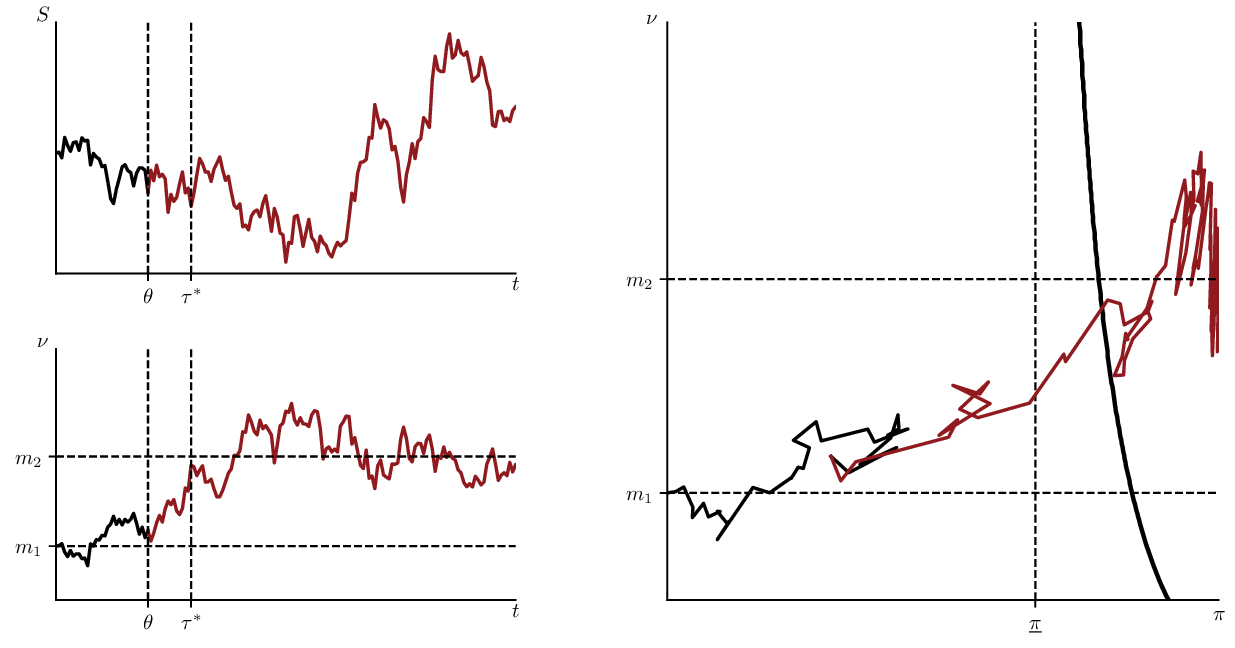}
    \caption{Optimal stopping boundary for the oracle version of the Heston model with a single change-point. Here $m_1<m_2$, depicted by horizontal dashed lines. The left panel is the asset price and variance process as time progresses while the two-dimensional diffusion $(\pi,\nu)$ is graphed in the right panel with the optimal stopping boundary (black curve) and $\underline{\pi}$ (dashed vertical line). The parameters are as in Section \ref{sec:num_heston} with $(S_0,\nu_0,Z_0,\mu,\kappa,\xi,\rho,\lambda)=(100,0.04,1,0.01,2,0.25,-0.7,2)$ and the change-point changing from $m_1=0.04$ to \(m_2=0.09\). The optimal stopping boundary and $\underline{\pi}$ are calculated under the quickest detection risk in \eqref{eq:qd_simple_risk} with $b=c=1$ and an infinite horizon.} %
    \label{fig:oracle_boundary}
\end{figure}

In the infinite-horizon quickest detection case with $m_1<m_2$, the value function is twice continuously differentiable in the interior of the continuation set and has continuous first derivatives across the interior free boundary, as discussed in Appendix \ref{app:oracle}. It therefore satisfies the following free-boundary problem in the classical sense %
\begin{align}
    \mathcal{L}V(\nu,\pi)&=-h(\pi) && (\nu,\pi)\in C \label{eq:fbp_first}\\
    V(\nu,\pi)&=g(\pi) && (\nu,\pi)\in D \\
    V_\pi(\nu,\pi)&=g'(\pi) && (\nu,\pi)\in\partial D\\
    V_\nu(\nu,\pi)&=0 && (\nu,\pi)\in\partial D \label{eq:fbp_last}
\end{align}
where $\mathcal{L}$ is the infinitesimal generator of $(\nu_t,\pi_t)$. The first equation follows from the dynamic programming principle in $C$, the second from immediate stopping in $D$, and the final two are the smooth-fit conditions at the interior free boundary. Furthermore, $C$, called the continuation set, and $D$, called the stopping set, form a partition of the state space of $(0,\infty)\times[0,1]$, with $\partial D$ being the boundary of $D$. 
In this case, the continuation and stopping sets can be characterized by
\begin{equation*}
    C = \{(\nu,\pi)\mid \pi<\beta(\nu)\}\quad\text{and} \quad D = \{(\nu,\pi)\mid \pi\geq \beta(\nu)\},
\end{equation*}
where $\beta:(0,\infty)\to(\underline{\pi},1)$ is a nonincreasing function, and $\underline{\pi}=b\lambda/(c+b\lambda)$. The optimal stopping time realizing the infimum in \eqref{eq:qd_heston_markov} is the first entry time to the stopping set $D$. The solution is plotted in the right panel of Figure \ref{fig:oracle_boundary} together with a simulated path of $(S_t,\nu_t)$ in the left panel for the quickest detection problem.

\subsection{Latent variance case}\label{sec:heston_filter}
We now return to the original problem in which the variance process $(\nu_t)$ is latent, so that $Y_t=S_t$, and the path $(\nu_t)$ is not directly observed.
The pressing issue is that the posterior probability process $(\pi_t)$ need not be a Markov process.

Indeed, for example, if $\lambda>0$ and $\pi_0<1$, then $(\pi_t)$ is not a Markov process by the following argument. Fix a $t>0$. Using the conditional second moment limit, similar to \eqref{eq:qvar_diff}, we have
\begin{equation}
    \underset{h\searrow0}{\mathrm{plim}}\frac{1}{h}\mathbb{E}[(\pi_{t+h}-\pi_t)^2\mid \mathcal{F}^\pi_t]=\eta_\rho(\nu_t)^2\pi_t^2(1-\pi_t)^2.
\end{equation}
However, if $(\pi_t)$ was Markov, then $\eta_\rho(\nu_t)^2\pi_t^2(1-\pi_t)^2$ should be a measurable function of $\pi_t$ (and possibly time). That is, there should exist a measurable function $\gamma:(0,\infty)\times(0,1)\to(0,\infty)$ such that $\gamma(t,\pi_t)$ equals $\nu_t$ almost surely. It is easy to see that no such function can exist for $\rho\neq0$. In the case of $\rho=0$, one may easily conclude that no such $\gamma$ exists in $C^{1,2}_\mathrm{loc}$ by Ito. To conclude no measurable $\gamma$ exists, one may note that the drift and diffusion coefficients of $(\nu_t)$ are analytic, and satisfy the no trap condition and so Hörmander's condition holds, see \cite[][proof of Theorem 6]{ernst2024gapeev}. This implies that $(\nu_t,\pi_t)$ has a two dimensional density for $t>0$. However, $\{(\nu,\pi)\mid \nu=\gamma(t,\pi)\}$ has two-dimensional Lebesgue measure zero for every $t>0$, and thus $\mathbb P(\nu_t=\gamma(t,\pi_t))=0$. So we reach a contradiction and conclude that $(\pi_t)$ is not a Markov process. %

A natural approach is to filter out an estimate of the $(\nu_t)$ process, e.g., $\mathbb{E}[\nu_t\mid\mathcal{F}^Y_t]$. However, in a continuous-time setting, this filter gives exactly $\nu_t$. %
Instead of considering only a point estimate of $\nu_t$ together with the posterior probability process, we consider the full conditional law $\Pi_t$ of $(\nu_t,Z_t)$ given $\mathcal F^Y_t$ and write its action on a test function as
\begin{equation}\label{eq:filter}
    \Pi_t(\varphi)=\mathbb{E}\left[\varphi(\nu_t,Z_t)\mid \mathcal{F}^Y_t\right],
\end{equation}
for measurable test-functions $\varphi:E\to\mathbb R$ satisfying $\mathbb E[|\varphi(\nu_t,Z_t)|]<\infty$, where $E=(0,\infty)\times\mathcal S$. Thus, $\Pi_t$ denotes the conditional probability measure in $\mathcal P(E)$, while $\Pi_t(\varphi)$ denotes its scalar evaluation at the test function $\varphi$. We equip $\mathcal S$ with the discrete topology and $\mathcal P(E)$ with the topology of weak convergence and its Borel sigma-field. %

Note we may write the posterior probability as
\begin{equation}\label{eq:heston_cl_tower}
    \Pi_t(1_{\{z=2\}})=\mathbb{P}(Z_t=2\mid\mathcal{F}^Y_t)
    =\mathbb{E}\left[
      \mathbb{P}(Z_t=2\mid\mathcal{F}^{S,\nu}_t)
      \mid\mathcal{F}^Y_t\right].
\end{equation}
The inner probability is the joint-observation posterior in \eqref{eq:posterior_heston_joint}. Under ideal continuous observation, \eqref{eq:qvar_diff} makes this identity redundant. Its discrete-observation analogue nevertheless still holds using the tower property, which motivates a conditional-likelihood (CL) approximation, see Section \ref{sec:num_heston}. %

It can be shown that, for test functions $\varphi$ in the domain of $\mathcal{L}$,
\begin{equation}\label{eq:ks_calender}
        {\rm d}\Pi_t(\varphi)=\Pi_t(\mathcal{L}\varphi){\rm d}t+\Pi_t(\sqrt{\nu})\Big(\Pi_t(\varphi f)-\Pi_t(\varphi)\Pi_t(f)+\rho\xi\,\Pi_t(\varphi_v)\Big){\rm d}I_t,
\end{equation}
where $(I_t)$ is a Brownian motion under $(\mathcal{F}^{Y}_t)$, see Appendix \ref{app:filters} for the details.
Equation \eqref{eq:ks_calender} is a measure-valued stochastic differential equation that forms a closed system.  %
The conditional distribution $(\Pi_t)$ defined by \eqref{eq:filter} exists and is unique up to versions and satisfies \eqref{eq:ks_calender}. Taking  $(\Pi_t)$ as being defined from \eqref{eq:ks_calender} requires additional conditions for well-posedness, and we will not pursue this in the current paper, see \cite{bain2009fundamentals}. %
It is shown in Appendix \ref{app:filters} that $(\Pi_t)$ is indeed a strong Markov process.

Thus, in order to solve the problem we can lift the original optimal stopping problem to a measure-valued optimal stopping problem and use the strong Markov property to get  
\begin{equation}\label{eq:Vtilde_heston}
    V(\Pi) = \inf_\tau \mathbb{E}\left[G(\Pi_\tau)+\int_0^\tau H(\Pi_u)\,{\rm d}u\,\Big\vert\, \Pi_0=\Pi\right],
\end{equation}
where $G,H:\mathcal P(E)\to[0,\infty)$ are measurable functionals defined by 
\begin{equation*}
    G(\Pi)=g\big(\Pi(1_{\{z=2\}})\big),\quad\text{and}\quad  H(\Pi)=h\big(\Pi(1_{\{z=2\}})\big),
\end{equation*}
where $g$ and $h$ were given in Section \ref{sec:heston_decision}. 
Hence, the optimal decision rule is given by some first hitting time in $\mathcal P(E)$. %
Representing the state by $\Pi_t$ gives a measure-valued Markov formulation. In the ideal continuous-observation Heston model, $\Pi_t$ is equivalently parameterized by $(\nu_t,\pi_t)$.
But retaining the measure-valued representation is useful as illustrated by the numerical example below, see Section \ref{sec:num_heston}.
Furthermore, in general regime switching multi-factor models, the oracle problem is only feasible when the latent variables are reasonably identifiable and the regime-switching model satisfies restrictive assumptions such as low number of factors, no feedback between the hidden regime and the factor model, etc. We will, however, see that lifting to a measure-valued optimal stopping problem can handle very general model settings.

\section{Regime switching multi-factor models}\label{sec:general_setup}
We now consider a general regime switching multi-factor model with a $d_Y$-dimensional observation process $(\bm{Y}_t)$ and a $d_X$-dimensional latent process $(\bm{X}_t)$. Let $(\Omega,\mathcal{F},(\mathcal{F}_t)_{t\geq0},\mathbb{P})$ be a complete %
filtered probability space with $\mathbb{P}$ being the physical measure. The space supports a $d_Y+d_X$-dimensional uncorrelated Brownian motion $(\bm{B}_t)$. We assume the observation process has dynamics 
\begin{align}\label{eq:obs_sde}
    {\rm d}\bm{Y}_t=\bm{\mu}(\bm{Y}_t,\bm{X}_t,Z_t)\,{\rm d}t+\mathbf{\Sigma}(\bm{Y}_t,\bm{X}_t,Z_t)\,{\rm d}\bm{B}_t,
\end{align}
where $\bm{\mu}:\mathbb{R}^{d_Y}\times\mathbb{R}^{d_X}\times\mathcal{S}\to \mathbb{R}^{d_Y}$, and $\mathbf{\Sigma}:\mathbb{R}^{d_Y}\times\mathbb{R}^{d_X}\times\mathcal{S}\to \mathbb{R}^{d_Y\times(d_Y+d_X)}$ are measurable %
and $\bm{Y}_0=y$. For simplicity, we assume that the first coordinate process of $(\bm{Y}_t)$ is $t$, the calendar time. The latent factors have dynamics
\begin{align}\label{eq:lat_sde}
    {\rm d}\bm{X}_t=\bm{\alpha}(\bm{Y}_t,\bm{X}_t,Z_t)\,{\rm d}t+\mathbf{\Xi}(\bm{Y}_t,\bm{X}_t,Z_t)\,{\rm d}\bm{B}_t,
\end{align}
where $\bm{\alpha}:\mathbb{R}^{d_Y}\times\mathbb{R}^{d_X}\times\mathcal{S}\to \mathbb{R}^{d_X}$, and $\mathbf{\Xi}:\mathbb{R}^{d_Y}\times\mathbb{R}^{d_X}\times\mathcal{S}\to \mathbb{R}^{d_X\times(d_Y+d_X)}$ are measurable. %
The hidden regimes are governed by the continuous-time Markov chain $(Z_t)$ with state space $\mathcal{S}=\{1,\dots,n\}$ which has intensity matrix $\mathbf{Q}:\mathbb{R}^{d_Y}\times\mathbb{R}^{d_X}\to \mathbb{R}^{n\times n}$. That is, for $\mathbf{Q}(\cdot)=\{q_{ij}(\cdot)\}_{i,j\in\mathcal{S}}$ it holds that the row sums vanish, i.e.,
\begin{equation*}
    -q_{ii}(\bm{y},\bm{x})=\sum_{j\in\mathcal{S}\setminus\{i\}} q_{ij}(\bm{y},\bm{x}),
\end{equation*}
for $(\bm{y},\bm{x})$ in the state space and all $i\in\mathcal{S}$.
We assume throughout that the model is well posed and that $(\bm Y_t,\bm X_t,Z_t)$ is a non-explosive strong Markov process. We denote the prior on the latent and hidden variables by $(\bm{X}_0,Z_0)\sim\Pi$. It is clear that the single-change-point decision Heston model considered in Section \ref{sec:heston} is a special case of this model. However, we allow for a very general dependence structure between the observations, the latent factors and the hidden regime.

\subsection{Decision-theoretic objectives}\label{sec:decision_gen}
A range of different decision-theoretic objectives can be considered. As an extension of the single-change-point problem considered in Section \ref{sec:heston_decision}, we now allow for the regime to switch multiple times. 
We will consider a non-empty subset of regimes $\mathcal{S}_\mathrm{1}\subset\mathcal{S}$ to be the normal states and the rest $\mathcal{S}_2=\mathcal{S}\setminus\mathcal{S}_1$ are considered abnormal regimes. %
We wish to detect when we are in an abnormal regime and declare which regime we are in. The decision rules are $(\tau,d)$ where $\tau$ is a $(\mathcal{F}^{\bm{Y}}_t)$-stopping time and 
$d\in\mathcal{S}_2$ %
is $\mathcal{F}^{\bm{Y}}_\tau$-measurable. Typical loss elements to consider are the probability of false alarm, expected detection delay, and a misclassification loss, say the risk 
\begin{equation}\label{eq:qd_general_risk}
    R(\tau,d)=\sum_{i\in \mathcal{S}_2}a_i\mathbb{P}(d=i,Z_\tau\neq i)+b\mathbb{P}(Z_\tau\in\mathcal{S}_1)+c\mathbb{E}\left[\int_0^\tau 1_{\{Z_t\in\mathcal{S}_2\}}\,{\rm d} t\right],
\end{equation}%
with constants $a_i,b,c>0$ for all $i\in\mathcal{S}_2$. 
Here the misclassification loss is an asymmetric ``0/1"-loss to account for different costs associated with wrongfully declaring the particular abnormal regimes. %
Minimizing over decision rules $(\tau,d)$, we get the Bayes risk
\begin{equation*}
    V=\inf_{(\tau,d)}R(\tau,d)=\inf_{\tau}\mathbb{E}\left[\bigwedge_{i\in\mathcal{S}_2}\left( a_i(1-\pi^{(i)}_\tau)\right)+b\sum_{i\in\mathcal{S}_1}\pi^{(i)}_\tau+ \int_0^\tau c\sum_{i\in\mathcal{S}_2}\pi^{(i)}_t\,{\rm d}t\right],
\end{equation*}%
where $\pi^{(i)}_t=\mathbb{P}(Z_t=i\mid \mathcal{F}^{\bm{Y}}_t)$ is the $i$th posterior probability process for $i\in\mathcal{S}$. The optimal decision is given by 
$d^*=\text{argmin}_{i\in\mathcal{S}_2}a_i(1-\pi^{(i)}_{\tau^*})$ %
where $\tau^*$ realizes the infimum in the optimal stopping problem. Similarly, one could consider more general loss functions for false alarm, detection delay, and misclassification, so for generality, we consider problems where the Bayes risk can be written as an optimal stopping problem of the form 
\begin{equation}\label{eq:general_one_decision}
    V = \inf_\tau \mathbb{E}\left[g\big(\tau,(\pi^{(i)}_\tau)_{i=1}^n\big)+\int_0^\tau h\big(t,(\pi^{(i)}_t)_{i=1}^n\big)\,{\rm d} t\right],
\end{equation}
where we also allow for time dependence and take $g,h:[0,\infty)\times\Delta_{n-1}\to\mathbb R$ to be measurable, where $\Delta_{n-1}=\{p\in[0,1]^n:\sum_{i=1}^n p_i=1\}$. %

\subsection{Measure-valued optimal stopping problem}
As discussed in Section \ref{sec:heston_filter}, the posterior probability processes $((\pi^{(i)}_t)_{i=1}^n)$ will in general not be Markov.
Similarly, $\big(\bm{Y}_t,(\pi^{(i)}_t)_{i=1}^n\big)$, or even adding the conditional expectations
$\hat{X}^{(i)}_t=\mathbb{E}[X^{(i)}_t\mid\mathcal{F}^{\bm Y}_t]$ to get $\big(\bm{Y}_t,(\pi^{(i)}_t)_{i=1}^n,
(\hat{X}^{(i)}_t)_{i=1}^{d_X}\big)$, may still fail to be Markov.

Instead, consider the full conditional law of $(\bm X_t,Z_t)$ given $\mathcal F^{\bm Y}_t$, denote this random probability measure by $\Pi_t$, and write its action on a test function as
\begin{equation*}
    \Pi_t(\varphi)
    =
    \mathbb{E}\left[
        \varphi(\bm X_t,Z_t)
        \mid\mathcal F^{\bm Y}_t
    \right],
\end{equation*}
for measurable test-functions $\varphi:\mathbb{R}^{d_X}\times\mathcal S\to\mathbb{R}$ for which the expectation is finite. Writing $E=\mathbb{R}^{d_X}\times\mathcal S$, $\Pi_t$ takes values in $\mathcal P(E)$, while $\Pi_t(\varphi)$ denotes its scalar evaluation at $\varphi$. The pair $(\bm Y_t,\Pi_t)$ is a strong Markov process. %
Indeed, let $\tau$ be an $(\mathcal F^{\bm Y}_t)$-stopping time and let $f$ be bounded and measurable. By the strong Markov property of $(\bm Y_t,\bm X_t,Z_t)$ and restarting the filtering problem from $\Pi_\tau$, for $s\geq0$ there exists a measurable function $K_s$ such that
\begin{equation*}
\mathbb{E}\left[
    f(\bm Y_{\tau+s},\Pi_{\tau+s})
    \mid\mathcal F^{\bm Y,\bm X,Z}_\tau
\right]
=
K_s(\bm Y_\tau,\bm X_\tau,Z_\tau,\Pi_\tau).
\end{equation*}
Taking conditional expectations with respect to
$\mathcal F^{\bm Y}_\tau$ gives
\begin{equation}\label{eq:filter_markov}
\mathbb{E}\left[
    f(\bm Y_{\tau+s},\Pi_{\tau+s})
    \mid\mathcal F^{\bm Y}_\tau
\right]
=
\int_E K_s(\bm Y_\tau,\bm x,z,\Pi_\tau)
       \Pi_\tau({\rm d}\bm x,{\rm d}z)=:Q_sf(\bm Y_\tau,\Pi_\tau),
\end{equation}
which proves the strong Markov property.
Thus, we can lift the optimal stopping problem in \eqref{eq:general_one_decision} and consider the optimal stopping problem 
\begin{equation}\label{eq:general_one_decision_lift}
    V = \inf_\tau \mathbb{E}\left[G\big(\tau,\Pi_\tau\big)+\int_0^\tau H\big(t,\Pi_t\big)\,{\rm d} t\right],
\end{equation}
with $(t,\bm Y_t,\Pi_t)$ as the state, and where $G,H:[0,\infty)\times\mathcal P(\mathbb R^{d_X}\times\mathcal S)\to\mathbb R$ are measurable functionals given by %
\begin{equation*}
    G(t,\Pi)=g\Big(t,\big(\Pi(1_{\{z=i\}})\big)_{i=1}^n\Big),\quad\text{and}\quad H(t,\Pi)=h\Big(t,\big(\Pi(1_{\{z=i\}})\big)_{i=1}^n\Big).
\end{equation*}
The use of the conditional law as a Markov state in partially observed optimal stopping is well established, see for example \citet{ludkovski2009simulation}.

\section{Numerical schemes}\label{sec:num_theory}
In the case when finite-dimensional Markovianity can be established, then it may be possible to prove that the optimal stopping problem \eqref{eq:general_one_decision} has enough regularity to be equivalent to a well-posed free-boundary problem like that of \eqref{eq:fbp_first}--\eqref{eq:fbp_last}, which can be solved by a standard PSOR algorithm \citep[see, e.g.,][]{duffy2013finite}. However, performance of finite-difference-based algorithms like the PSOR drastically deteriorates with dimensionality, and furthermore, boundary regularity for non-monotone optimal stopping problems, to a large extent, %
requires ad hoc analysis. Finally, in applications where the finite-dimensional Markovianity can be established through identifiability of the latent process, like in Section \ref{sec:heston_oracle}, then performance hinges on the approximation through the identifiability link, e.g., through the realized variance approximation \eqref{eq:rvar} for stochastic volatility models.

As an alternative, we can consider the measure-valued optimal stopping problem in \eqref{eq:general_one_decision_lift}. For tractability, we consider the problem on an equidistant discrete-time grid with time step $\Delta$ and a finite time horizon $T$. The factor processes evolve by 
\begin{equation}\label{eq:obs_rec}
    \bm{Y}_{t+\Delta}=\bm{f_Y}(\bm{Y}_t,\bm{X}_{t+\Delta},Z_{t+\Delta},\bm{N}_{t+\Delta},\Delta)
\end{equation}
and 
\begin{equation}\label{eq:lat_rec}
    \bm{X}_{t+\Delta}=\bm{f_X}(\bm{Y}_t,\bm{X}_t,Z_t,\bm{N}_{t+\Delta},\Delta)
\end{equation}
for $\bm{N}_{t+\Delta}$ being i.i.d. (Gaussian) noise, and $\bm{f_Y}$ and $\bm{f_X}$ being appropriate approximation schemes of \eqref{eq:obs_sde} and \eqref{eq:lat_sde}, e.g. an Euler-Maruyama scheme. 
We let $(Z_t)$ be a discrete time Markov chain with transition matrix 
\begin{equation}\label{eq:P_from_Q}
    \mathbf{P}(t)=e^{\Delta \mathbf{Q}(\bm{Y}_t,\bm{X}_t)}, 
\end{equation}
and
\begin{equation}\label{eq:hid_rec}
    \mathbb{P}(Z_{t+\Delta}=j\mid Z_t=i)=(\mathbf{P}(t))_{ij},
\end{equation}
which can be done by a standard transition scheme.
Under standard regularity conditions, Euler-type approximations of state-dependent regime-switching diffusions have discretization error of order \(O_p(\Delta^{1/2})\), see, e.g., \citet{shao2018invariant}. %
In both numerical experiments below, the regime transition probabilities are sampled exactly on the observation grid, so this additional switching-discretization error is absent.

Discretizing \eqref{eq:general_one_decision_lift} and using the Markov property of $(\bm Y_t,\Pi_t)$ we consider the discrete time Markovian optimal stopping problem 
\begin{equation*}
    V(t,\bm{Y},\Pi) = \inf_{\tau\leq T-t} \mathbb{E}\Bigg[G\big(t+\tau,\Pi_{t+\tau}\big)+\sum_{i=0}^{\tau/\Delta-1} H\big(t+i\Delta,\Pi_{t+i\Delta}\big)\Delta\,\Bigg\vert\,\bm{Y}_t=\bm{Y},\Pi_t=\Pi\Bigg], %
\end{equation*}
where the infimum is taken over discrete stopping times with $\tau\leq T-t$. By the dynamic programming principle, we can write 
\begin{equation}\label{eq:disc_osp_markov}
    V(t,\bm Y, \Pi) = \min\left\{G\big(t,\Pi\big),H(t,\Pi)\Delta+\mathbb{E}[V(t+\Delta,\bm Y_{t+\Delta},\Pi_{t+\Delta})\,\big\vert\,\bm Y_t=\bm Y,\Pi_t=\Pi\big]\right\},
\end{equation}
due to the Markovianity.

\subsubsection*{Particle-filter Monte Carlo dynamic programming.} %
To numerically approximate this, we use a PFMCDP implementation, following the particle-filter/regression Monte Carlo approach of \citet{ludkovski2009simulation,ludkovski2012bayesian}.
Given simulated filtered states, Monte Carlo dynamic programming approximates the conditional expectation in \eqref{eq:disc_osp_markov} by a regression
\begin{equation}\label{eq:CV_regression}
    \begin{aligned}
    \mathrm{CV}_K(t,\bm Y,\Pi)&=\bm\beta_Y(t)^\top\bm Y+\sum_{k=1}^K \beta_k(t)B_k(\Pi)
    \\&\approx\mathbb{E}\left[V(t+\Delta,\bm Y_{t+\Delta},\Pi_{t+\Delta})\mid \bm Y_t=\bm Y,\Pi_t=\Pi\right],
    \end{aligned}
\end{equation}
where each $B_k:\mathcal P(E)\to\mathbb R$ is a regression feature functional of the filtered measure. In particular, $B_k(\Pi)$ may be a linear evaluation $\Pi(\varphi)$ or a nonlinear transformation or cross term of such filtered summaries. %

To simulate the filtered states, we approximate $\Pi$ by a weighted empirical measure using a particle filter. Appendix \ref{app:pf} gives the bootstrap particle filter and the exact regression covariates used in the numerical experiments.  %
In an offline phase, we simulate the filtered states and calculate the regression coefficients recursively backwards with $V(T,\bm Y,\Pi)=G(T,\Pi)$. Then in the online phase, as observations $\bm{Y}_t$ arrive sequentially, we estimate the measure by a particle filter $\hat{\Pi}_t$ and evaluate the stopping rule by
\begin{equation*}
    \tau= \inf\{t=0,\Delta,\dots,T\mid G(t,\hat{\Pi}_t)\leq H(t,\hat{\Pi}_t)\Delta+\mathrm{CV}_K(t,\bm Y_t,\hat{\Pi}_t)\}\wedge T,
\end{equation*}
and the decision rule is taken in accordance with the particular risk considered through the discussion in Section \ref{sec:decision_gen}.

\begin{remark} %
    It is important to note that the individual parts of the PFMCDP algorithm can easily be replaced by more advanced methods. For instance, one may use modern particle filters such as \cite{gellert2021parameter}, which studies genetic-algorithm-inspired kernel-smoothing adaptations designed to improve performance under regime changes. For very high-dimensional models, it is natural to replace the regression step by a neural network as in \cite{becker2021solving} and to consider differentiable particle filters as proposed by \cite{brady2026differentiable}. One may instead prefer Kalman-type filtering methods for computational speed when suitable approximations are available. However, for Heston-type stochastic volatility models, \cite{LI201315} finds that stock-price observations alone may be insufficient for accurate volatility estimation in an unscented Kalman filter, while in regime-switching models, Kalman-based methods require hypothesis reduction to control the growth in possible regime histories \cite{1299}.
\end{remark} %

\section{Numerical experiments}\label{sec:num}
We report two numerical experiments. The first considers a regime-switching Vasicek model with measurement-noise-contaminated bond prices and is used to assess the sensitivity of the PFMCDP procedure to its main approximation parameters. The second revisits the single-change-point Heston model from Section \ref{sec:heston}, where the partially observed procedure is benchmarked against the oracle problem.

\subsection{Measurement-noise example}
We consider a regime-switching Vasicek model in which the latent short rate $(X_t)$ evolves under the physical measure $\mathbb{P}$ as
\begin{equation*}
    {\rm d}X_t=\kappa(\theta-X_t){\rm d}t+\sigma {\rm d}B_t,
\end{equation*}
where $\kappa,\sigma>0$ and $\theta\in\mathbb{R}$. The market price of interest-rate risk is assumed regime dependent and constant within each regime:
\begin{equation*}
    {\rm d}B_t^{\mathbb Q} = {\rm d}B_t + \zeta_{Z_t}\,{\rm d}t,%
\end{equation*}
where $(Z_t)$ is the continuous-time Markov chain governing the regime, $\zeta_i\in\mathbb R$ %
for $i\in\mathcal S$, and $B^{\mathbb Q}$ is a Brownian motion under the pricing measure $\mathbb Q$. 
Let $p:[0,T]\times\mathbb R\times \mathcal{S}\to \mathbb R_+$ denote the measurable zero-coupon bond-price function (see Appendix \ref{app:vasicek}). %
Observations are generated from
\begin{equation}\label{eq:obs_eq_noise_revised}
    Y_t = p(t,X_t,Z_t)+\varepsilon_t,
\end{equation}
where $(\varepsilon_t)$ is an i.i.d. centered Gaussian measurement-noise sequence with variance $\xi^2$. %
Hence, the observation equation is already of the form \eqref{eq:obs_rec}. In the numerical implementation, the Markov chain is sampled on the observation grid by the transition matrix $\mathbf{P}$ given in \eqref{eq:P_from_Q} and simulated by \eqref{eq:hid_rec},
while the latent short rate is discretized by the Euler--Maruyama scheme.

We consider a one-year horizon $T=1$ with parameters
\[
(\kappa,\theta,\sigma,\xi)=(1.5,0.05,0.02,0.005).
\]
The regime process has three states, $(\zeta_i)_{i\in\mathcal{S}}=(0,3,-1/2)$ corresponding to a neutral regime,
$\mathcal{S}_1=\{1\}$, %
and two abnormal regimes,
$\mathcal{S}_2=\{2,3\}$, %
of greed (willing to pay more for risk) and fear (needing more compensation for taking risk), respectively. The generator is
\[
    \mathbf{Q}= \begin{bmatrix}
        -0.75& 0.25 &   0.5\\
          2.0& -2.0 &   0.0\\
          1.0& 0.25 & -1.25
    \end{bmatrix}.
\]
Misclassification losses are asymmetric, while we take false-alarm and detection-delay weights to be equal, $(a_2,a_3,b,c)=(0.2,0.6,0.8,0.8)$. The latent factor $X_0$ is initialized from its stationary distribution and the regime is initialized in the known normal state $Z_0=1$. 
The parameters are chosen for a stylized numerical experiment rather than calibrated to a particular data set. They produce economically distinct regimes while keeping the filtering and detection problem nontrivial. %

For the PFMCDP algorithm, we use daily observations, $\Delta=1/250$, and a training/testing split of $5000/3000$ simulated paths. The baseline specification uses $M=500$ particles, resampling threshold $\alpha=0.5$, and the $K=15$ covariates described in Appendix \ref{app:pf}. Furthermore, the covariates are standardized, and we perform a ridge regression with two-fold cross-fitting to reduce overfitting. 
Table \ref{tab:sensitivity_results} reports the mean loss and standard error computed from $N_{\mathrm{rep}}=20$ independent replications. Specifically, for replication $r=1,\ldots,N_{\mathrm{rep}}$, we calculate the empirical risk over $N_{\mathrm{test}}$ test paths as
\begin{equation}\label{eq:empirical_risk}
\hat R_r =
\frac{1}{N_{\mathrm{test}}}
\sum_{i=1}^{N_{\mathrm{test}}}
\left(
\sum_{j\in\mathcal S_2} a_j \mathbf 1_{\{d_{r,i}=j,\, Z^{(r,i)}_{\tau_{r,i}}\neq j\}}
+
b\mathbf 1_{\{Z^{(r,i)}_{\tau_{r,i}}=1\}}
+ 
c\Delta \sum_{k=0}^{\tau_{r,i}/\Delta-1} \mathbf 1_{\{Z^{(r,i)}_{k\Delta}\neq1\}} 
\right),
\end{equation}
where $(\tau_{r,i},d_{r,i})$ is the decision rule trained in replication $r$ and applied to the $i$th test path of that replication. %

\begin{table}[htbp]
\centering
\caption{
Expected loss using PFMCDP for a range of parameters. Reported values are the sample mean and standard error (in parentheses) of the empirical risks $(\hat R_r)_{r=1}^{N_{\mathrm{rep}}}$ defined in \eqref{eq:empirical_risk}, with $N_{\mathrm{rep}}=20$ independent training-and-testing replications. %
}
\label{tab:sensitivity_results}
\begin{tabular}{lccccc}
\toprule
\multicolumn{6}{c}{\textbf{Panel A: Sensitivity with respect to $M$ and $\alpha$}} \\
\midrule
& $\alpha = 0.3$ & $\alpha = 0.4$ & $\alpha = 0.5$ & $\alpha = 0.6$ & $\alpha = 0.7$ \\
\midrule
$M=50$
& \begin{tabular}[c]{@{}c@{}}0.8631\\(0.0011)\end{tabular}
& \begin{tabular}[c]{@{}c@{}}0.8605\\(0.0009)\end{tabular}
& \begin{tabular}[c]{@{}c@{}}0.8611\\(0.0011)\end{tabular}
& \begin{tabular}[c]{@{}c@{}}0.8602\\(0.0013)\end{tabular}
& \begin{tabular}[c]{@{}c@{}}0.8607\\(0.0010)\end{tabular} \\[0.3em]
$M=100$
& \begin{tabular}[c]{@{}c@{}}0.8532\\(0.0010)\end{tabular}
& \begin{tabular}[c]{@{}c@{}}0.8505\\(0.0013)\end{tabular}
& \begin{tabular}[c]{@{}c@{}}0.8496\\(0.0015)\end{tabular}
& \begin{tabular}[c]{@{}c@{}}0.8482\\(0.0014)\end{tabular}
& \begin{tabular}[c]{@{}c@{}}0.8482\\(0.0013)\end{tabular} \\[0.3em]
$M=250$
& \begin{tabular}[c]{@{}c@{}}0.8367\\(0.0011)\end{tabular}
& \begin{tabular}[c]{@{}c@{}}0.8344\\(0.0008)\end{tabular}
& \begin{tabular}[c]{@{}c@{}}0.8301\\(0.0011)\end{tabular}
& \begin{tabular}[c]{@{}c@{}}0.8301\\(0.0012)\end{tabular}
& \begin{tabular}[c]{@{}c@{}}0.8307\\(0.0014)\end{tabular} \\[0.3em]
$M=500$
& \begin{tabular}[c]{@{}c@{}}0.8243\\(0.0012)\end{tabular}
& \begin{tabular}[c]{@{}c@{}}0.8224\\(0.0012)\end{tabular}
& \begin{tabular}[c]{@{}c@{}}0.8202\\(0.0015)\end{tabular}
& \begin{tabular}[c]{@{}c@{}}0.8213\\(0.0014)\end{tabular}
& \begin{tabular}[c]{@{}c@{}}0.8200\\(0.0011)\end{tabular} \\[0.3em]
$M=1000$
& \begin{tabular}[c]{@{}c@{}}0.8162\\(0.0013)\end{tabular}
& \begin{tabular}[c]{@{}c@{}}0.8156\\(0.0011)\end{tabular}
& \begin{tabular}[c]{@{}c@{}}0.8171\\(0.0013)\end{tabular}
& \begin{tabular}[c]{@{}c@{}}0.8146\\(0.0016)\end{tabular}
& \begin{tabular}[c]{@{}c@{}}0.8152\\(0.0015)\end{tabular} \\
\midrule
\multicolumn{6}{c}{\textbf{Panel B: Sensitivity with respect to $K$}} \\
\midrule
& $K=3$ & $K=4$ & $K=5$ & $K=9$ & $K=15$ \\
\midrule
Result
& \begin{tabular}[c]{@{}c@{}}0.8290\\(0.0014)\end{tabular}
& \begin{tabular}[c]{@{}c@{}}0.8277\\(0.0012)\end{tabular}
& \begin{tabular}[c]{@{}c@{}}0.8272\\(0.0011)\end{tabular}
& \begin{tabular}[c]{@{}c@{}}0.8232\\(0.0014)\end{tabular}
& \begin{tabular}[c]{@{}c@{}}0.8202\\(0.0015)\end{tabular} \\
\bottomrule
\end{tabular}
\end{table}

Panel A shows that the expected loss decreases materially as the particle count is increased to $M=500$. At $\alpha=0.5$, the mean loss falls from $0.8301$ for $M=250$ to $0.8202$ for $M=500$, while the further reduction to $0.8171$ for $M=1000$ is small relative to the replication uncertainty. The $250$--$500$ improvement is statistically significant, whereas the $500$--$1000$ difference is not significant at the 5\% level. At $M=500$, the results for $\alpha\in[0.4,0.7]$ are very similar, while $\alpha=0.3$ performs somewhat worse.

Panel B examines the nested filtered-state covariate sets $K=3,4,5,9,15$ defined in Appendix \ref{app:pf}. Mean loss decreases from $0.8290$ at $K=3$ to $0.8202$ at $K=15$. The main improvements arise when adding the regime-weighted moment summaries and the final nonlinear and interaction covariates, while adding the global variance alone gives only a small change.
The sensitivity analysis therefore supports $M=500$ and $\alpha=0.5$ as reasonable computational choices and the full set of covariates (up to a change in number of regimes). %

\subsection{Single-change-point Heston model}\label{sec:num_heston}
Next, we consider the quickest detection problem for the single-change-point Heston model introduced in Section \ref{sec:heston}. The aim is to compare the PFMCDP procedure with the oracle rule and with several natural plug-in benchmarks.

We consider a horizon of \(T=3\) years with parameters
\[
(S_0,\mu,\kappa,\xi,\rho)=(100,0.01,2,0.25,-0.7),
\]
and a change in the long-run variance level from \(m_1=0.04\) to \(m_2=0.09\). The change-point is exponentially distributed with hazard rate \(\lambda=2\). The regime process starts in the pre-change state, \(Z_0=1\). Both the true initial variance and the variance particles are initialized from the stationary Gamma distribution of the CIR dynamics in Regime \(1\). %
Similar to the measurement-noise example above, the parameters here are illustrative rather than calibrated. The shift from $m_1$ to $m_2$ produces a persistent increase in long-run variance, while the moderate speed of mean reversion and diffusion in $(\nu_t)$ keep the detection problem non-trivial, as illustrated in the left panel of Figure \ref{fig:oracle_boundary}. %

\begin{figure}[h]

\centering
\includegraphics[width=\linewidth]{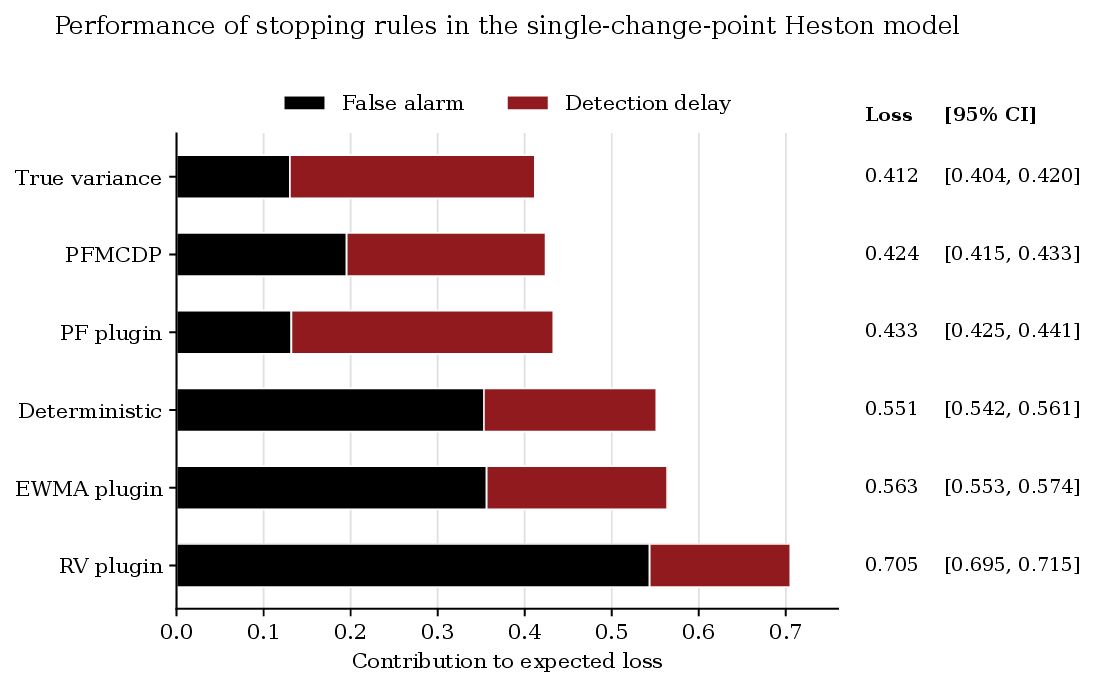} %
\caption{Performance of stopping rules in the single-change-point Heston model. Each bar decomposes the estimated risk in \eqref{eq:qd_simple_risk} into its false-alarm and detection-delay contributions, $b\widehat{\mathrm{PFA}}$ and $c\widehat{\mathrm{EDD}}$, with $b=c=1$. Total estimated losses and pointwise 95\% Monte Carlo confidence intervals ($N_\mathrm{test}=5000$), calculated from the pathwise losses in \eqref{eq:FA_DD} conditional on the fitted rules, are reported on the right. True variance denotes the oracle using both stock and variance paths. PF plugin is the particle-filter plug-in rule with the posterior \eqref{eq:heston_cl_posterior}. RV and EWMA are proxy rules and Deterministic is the best deterministic stopping rule.} %
\label{fig:heston_loss}
\end{figure}

The model is simulated on a fine observation grid with $\Delta = 1/(250\cdot 12)$ %
corresponding to approximately 12 intraday observations. We use a full-truncation Euler scheme of the form
\begin{align}
    Z_{t+\Delta} &= 1+\mathbf{1}_{\{Z_t=2\}}+\mathbf{1}_{\{Z_t=1,\;U_{t+\Delta}<1-e^{-\lambda\Delta}\}},\\
    \nu_{t+\Delta} &= \nu_t+\kappa\bigl(m_{Z_{t+\Delta}}-\nu_t^+\bigr)\Delta
    +\xi\sqrt{\nu_t^+\Delta}\,N^{(1)}_{t+\Delta},\\
    \log S_{t+\Delta} &= \log S_t+\Bigl(\mu-\frac12 \nu_{t+\Delta}^+\Bigr)\Delta
    +\sqrt{\nu_{t+\Delta}^+\Delta}\,N^{(2)}_{t+\Delta},
\end{align}
where \(U_{t+\Delta}\) is uniformly distributed on \([0,1]\), and $(N^{(1)}_{t+\Delta}, N^{(2)}_{t+\Delta})$ are standard Gaussian innovations with correlation \(\rho\).
For the PFMCDP method, we use $M=500$ particles and a training/testing split of $8000/5000$ simulated paths. The regression uses the $K=11$ covariates described in Appendix \ref{app:pf}, together with ridge regression and two-fold cross-fitting. Figure \ref{fig:heston_loss} reports the total loss together with its two components, the expected detection delay (EDD) and the probability of false alarm (PFA), for the symmetric choice \(b=c=1\).

For this single-change-point example, the direct particle estimate $\sum_{m=1}^M w_t^{(m)}\mathbf{1}_{\{Z_t^{(m)}=2\}}$ is vulnerable to regime-particle impoverishment. Once resampling removes every pre-change particle, the absorbing dynamics prevent their reintroduction, and the estimate remains equal to one. We therefore use the CL construction motivated by \eqref{eq:heston_cl_tower}.
For $t_k=k\Delta$, let $q_{t_k}^{(m)}$ be the joint-observation posterior \eqref{eq:posterior_heston_joint}, with \eqref{eq:likelihood_heston_joint} discretized along the observed stock path and the $m$th particle's variance path. We set
\begin{equation}\label{eq:heston_cl_posterior}
    \widehat\pi_{t_k}^{\mathrm{CL}}
    =\sum_{m=1}^M w_{t_k}^{(m)}q_{t_k}^{(m)}.
\end{equation}
This approximates the corresponding discrete-observation conditional construction rather than giving an exact filter on a fixed grid. The likelihood statistics $q_{t_k}^{(m)}$ follow particle ancestry through the resampling step of the bootstrap particle filter. We use $\widehat\pi^{\mathrm{CL}}$ and $1-\widehat\pi^{\mathrm{CL}}$ for the posterior-probability inputs to PFMCDP, including its stopping and running costs, in both training and testing. Particle propagation and weights remain as in Appendix \ref{app:pf}.

As a benchmark, we first solve the oracle problem, where both stock and variance paths are observed, by PSOR on a \(351\times 351\) grid and obtain the optimal stopping boundary \(\beta(\cdot)\). 
We use piecewise-linear interpolation of the computed boundary to determine first-passage times of $(\nu_t,\pi_t)$, with $\pi_t$ computed from the observed stock and true variance paths through the discretized version of \eqref{eq:posterior_heston_joint}--\eqref{eq:likelihood_heston_joint}. Although the resulting interpolant is non-smooth, it preserves the monotonicity of the computed boundary. Using spline interpolation instead has only a negligible effect on the numerical results reported below. %
We also report several feasible proxy rules obtained by replacing the unobserved variance in the oracle policy by an estimate. Specifically, we consider realized variance (RV), an exponentially weighted moving average (EWMA), and the particle-filter point estimate
\[
\widehat{\nu}_t=\sum_{i=1}^M w_t^{(i)}\nu_t^{(i)}.
\]
For RV and EWMA, the posterior is obtained by inserting the corresponding variance proxy and the observed stock path into the discretized formulas \eqref{eq:posterior_heston_joint}--\eqref{eq:likelihood_heston_joint}. The particle-filter plug-in rule, labelled ``PF plugin'', instead uses the state $(\widehat\nu_t,\widehat\pi_t^{\mathrm{CL}})$ and stops when $\widehat\pi_t^{\mathrm{CL}}\geq\beta(\widehat\nu_t)$, or at the terminal time $T$. Thus, it shares its posterior-probability estimate with PFMCDP rather than obtaining that coordinate by substituting $\widehat\nu$ into the likelihood.
Finally, we include the optimal deterministic stopping rule, that is, the best stopping rule among all pre-specified calendar times. Although the timing is non-adaptive, this provides a useful benchmark for assessing whether the additional complexity of observation-based proxy rules translates into an improved decision rule. The RV and EWMA windows, and the deterministic stopping time are calibrated on the training sample. %

For method $m$ and test path $i$, define the pathwise false-alarm, detection-delay, and quickest-detection losses by
\begin{equation}\label{eq:FA_DD}
L_{\mathrm{FA},i}^{(m)}=\mathbf 1_{\{\tau_i^{(m)}<\theta_i\}},
\qquad
L_{\mathrm{DD},i}^{(m)}=(\tau_i^{(m)}-\theta_i)^+,
\qquad
L_i^{(m)}=bL_{\mathrm{FA},i}^{(m)}+cL_{\mathrm{DD},i}^{(m)}
\end{equation}
where $\theta_i$ is the change point of the $i$th test path and
$\tau_i^{(m)}$ is the stopping time produced by method $m$ on that path.
The empirical risk and its components are then
\begin{equation}\label{eq:PFA_EDD_R}
\widehat{\mathrm{PFA}}^{(m)}
=
\frac1{N_{\mathrm{test}}}
\sum_{i=1}^{N_{\mathrm{test}}}
L_{\mathrm{FA},i}^{(m)},
\qquad
\widehat{\mathrm{EDD}}^{(m)}
=
\frac1{N_{\mathrm{test}}}
\sum_{i=1}^{N_{\mathrm{test}}}
L_{\mathrm{DD},i}^{(m)},
\qquad
\widehat R^{(m)}
=
\frac1{N_{\mathrm{test}}}
\sum_{i=1}^{N_{\mathrm{test}}}L_i^{(m)}.
\end{equation}
All methods are evaluated on the same test paths, independently of training. Figure \ref{fig:heston_loss} shows the estimated losses and their decomposition, with approximate 95\% confidence intervals for total loss calculated from the sample standard error of $(L_i^{(m)})_{i=1}^{N_\mathrm{test}}$ for each method. These intervals quantify test-sample Monte Carlo uncertainty conditional on the fitted decision rules and do not include variation from repeating the training procedure. %

Figure \ref{fig:heston_loss} shows that the oracle rule based on the observed stock and true variance attains the lowest estimated loss. Among the implementable rules, PFMCDP has the lowest estimated loss. Both particle-filter-based estimated losses are below those of RV, EWMA, and the deterministic benchmark. The decomposition also shows that RV and EWMA attain shorter detection delays at the expense of higher false-alarm rates compared with the true-variance and PF plug-in rules, even though they use the same stopping boundary. Notably, in this experiment the EWMA- and RV-based rules are also outperformed by the deterministic benchmark, illustrating that incorporating a noisy variance proxy into the oracle policy need not improve upon a simple non-adaptive timing rule. %

A particularly important comparison is between PFMCDP and the particle-filter plug-in rule. Both procedures use the same filtering output and conditional-likelihood posterior, but the plug-in rule inserts $(\widehat\nu_t,\widehat\pi_t^{\mathrm{CL}})$ into the oracle stopping rule, whereas PFMCDP optimizes the stopping decision directly under latent and hidden information. For the paired pathwise loss $D_i=L_i^{(\mathrm{PFMCDP})}-L_i^{(\mathrm{PF\ plugin})}$, $i=1,\ldots,5000$, the estimated mean is $-0.008813$ with 95\% Student-$t$ confidence interval $[-0.015252,\,-0.002374]$. A two-sided paired $t$-test of zero mean difference gives $p=0.007319$. Thus, the difference is statistically significant at the conventional 5\% level and favors PFMCDP. This comparison is conditional on the fitted rules and does not include training variability. Unlike the plug-in oracle construction, PFMCDP also applies to the general latent-factor setting without requiring a finite-dimensional oracle boundary. %

\section{Conclusion}
We have shown that the PFMCDP methodology can be applied to regime-switching multi-factor models to approximate optimal decision rules for regime detection in a decision-theoretic framework. The approach is flexible and can be combined with more advanced particle-filtering and dynamic-programming methods, for instance by using differentiable particle filters together with neural-network-based continuation-value approximations \citep{becker2021solving,brady2026differentiable}.

The framework also admits several natural extensions. One direction is to consider multiple stopping decisions, similar to swing-option problems in energy markets. Another is to study the actions taken after detection and incorporate feedback controls into the state dynamics.

\section*{Funding}
This research was partly supported by the travel scholarship Rejselegat for Matematikere.

\section*{Disclosure statement}
The author declares no conflicts of interest. The author alone is responsible for the content and writing of the paper.

\section*{Data availability statement}
The numerical results in this study are based on simulated data generated from the models described in the manuscript. No external datasets were used. The simulated datasets are available upon reasonable request.

\section*{Generative AI disclosure}
OpenAI's ChatGPT (GPT-6 Astra) was used to assist with language editing, reviewing mathematical derivations for potential inconsistencies, and checking LaTeX and bibliographic formatting. The tool was used to improve clarity and consistency. All suggested changes were reviewed by the author, who takes full responsibility for the manuscript.

\bibliographystyle{abbrvnat}
\bibliography{refs}

\begin{thebibliography}{58}
\providecommand{\natexlab}[1]{#1}
\providecommand{\url}[1]{\texttt{#1}}
\expandafter\ifx\csname urlstyle\endcsname\relax
  \providecommand{\doi}[1]{doi: #1}\else
  \providecommand{\doi}{doi: \begingroup \urlstyle{rm}\Url}\fi

\bibitem[Aihara and Bagchi(2003)]{AIHARA20031681}
S.~Aihara and A.~Bagchi.
\newblock Filtering of stochastic volatility model.
\newblock \emph{IFAC Proceedings Volumes}, 36\penalty0 (16):\penalty0 1681--1686, 2003.
\newblock 13th IFAC Symposium on System Identification (SYSID 2003), Rotterdam, The Netherlands, 27-29 August, 2003.

\bibitem[Babbs and Nowman(1999)]{babbs1999kalman}
S.~H. Babbs and K.~B. Nowman.
\newblock Kalman filtering of generalized vasicek term structure models.
\newblock \emph{Journal of Financial and Quantitative Analysis}, 34\penalty0 (1):\penalty0 115--130, 1999.

\bibitem[Bain and Crisan(2009)]{bain2009fundamentals}
A.~Bain and D.~Crisan.
\newblock \emph{Fundamentals of Stochastic Filtering}.
\newblock Stochastic Modelling and Applied Probability, Vol. 60. Springer, New York, 2009.

\bibitem[Baras and LaVigna(1987)]{baras1987real}
J.~S. Baras and A.~LaVigna.
\newblock Real-time sequential detection for diffusion signals.
\newblock In \emph{26th IEEE Conference on Decision and Control}, volume~26, pages 1153--1157. IEEE, 1987.

\bibitem[Becker et~al.(2021)Becker, Cheridito, Jentzen, and Welti]{becker2021solving}
S.~Becker, P.~Cheridito, A.~Jentzen, and T.~Welti.
\newblock Solving high-dimensional optimal stopping problems using deep learning.
\newblock \emph{European Journal of Applied Mathematics}, 32\penalty0 (3):\penalty0 470--514, 2021.

\bibitem[Blom and Bar-Shalom(1988)]{1299}
H.~Blom and Y.~Bar-Shalom.
\newblock The interacting multiple model algorithm for systems with markovian switching coefficients.
\newblock \emph{IEEE Transactions on Automatic Control}, 33\penalty0 (8):\penalty0 780--783, 1988.

\bibitem[Brady et~al.(2026)Brady, Luo, Wang, Elvira, and Li]{brady2026differentiable}
J.-J. Brady, Y.~Luo, W.~Wang, V.~Elvira, and Y.~Li.
\newblock Differentiable interacting multiple model particle filtering.
\newblock \emph{Signal Processing}, 238:\penalty0 110166, 2026.

\bibitem[Carr and Madan(1999)]{carr1999option}
P.~Carr and D.~Madan.
\newblock Option valuation using the fast fourier transform.
\newblock \emph{Journal of Computational Finance}, 2\penalty0 (4):\penalty0 61--73, 1999.

\bibitem[Carvalho and Lopes(2007)]{carvalho2007simulation}
C.~M. Carvalho and H.~F. Lopes.
\newblock Simulation-based sequential analysis of markov switching stochastic volatility models.
\newblock \emph{Computational Statistics \& Data Analysis}, 51\penalty0 (9):\penalty0 4526--4542, 2007.

\bibitem[Cheridito et~al.(2007)Cheridito, Filipovi{\'c}, and Kimmel]{cheridito2007market}
P.~Cheridito, D.~Filipovi{\'c}, and R.~L. Kimmel.
\newblock Market price of risk specifications for affine models: Theory and evidence.
\newblock \emph{Journal of Financial Economics}, 83\penalty0 (1):\penalty0 123--170, 2007.

\bibitem[Chiarolla and De~Angelis(2016)]{chiarolla2016optimal}
M.~B. Chiarolla and T.~De~Angelis.
\newblock Optimal stopping of a hilbert space valued diffusion: an infinite dimensional variational inequality.
\newblock \emph{Applied Mathematics \& Optimization}, 73\penalty0 (2):\penalty0 271--312, 2016.

\bibitem[Christensen(2013)]{christensen2013regime}
J.~H.~E. Christensen.
\newblock A regime-switching model of the yield curve at the zero bound.
\newblock Working Paper Series 2013-34, Federal Reserve Bank of San Francisco, 2013.

\bibitem[Collin-Dufresne and Fos(2015)]{collin2015prices}
P.~Collin-Dufresne and V.~Fos.
\newblock Do prices reveal the presence of informed trading?
\newblock \emph{The Journal of Finance}, 70\penalty0 (4):\penalty0 1555--1582, 2015.

\bibitem[Dai and Singleton(2000)]{dai2000specification}
Q.~Dai and K.~J. Singleton.
\newblock Specification analysis of affine term structure models.
\newblock \emph{The Journal of Finance}, 55\penalty0 (5):\penalty0 1943--1978, 2000.

\bibitem[De~Angelis and Peskir(2020)]{de2020global}
T.~De~Angelis and G.~Peskir.
\newblock Global {$C^1$} regularity of the value function in optimal stopping problems.
\newblock \emph{The Annals of Applied Probability}, 30\penalty0 (3):\penalty0 1007--1031, 2020.

\bibitem[Doucet and Johansen(2011)]{doucet2011tutorialPF}
A.~Doucet and A.~M. Johansen.
\newblock A tutorial on particle filtering and smoothing: Fifteen years later.
\newblock In D.~Crisan and B.~Rozovskii, editors, \emph{The Oxford Handbook of Nonlinear Filtering}, pages 656--704. Oxford University Press, Oxford, 2011.

\bibitem[Duan et~al.(2002)Duan, Popova, and Ritchken]{duan2002option}
J.-C. Duan, I.~Popova, and P.~Ritchken.
\newblock Option pricing under regime switching.
\newblock \emph{Quantitative Finance}, 2\penalty0 (2):\penalty0 116--132, 2002.

\bibitem[Duffie and Kan(1996)]{duffie1996yield}
D.~Duffie and R.~Kan.
\newblock A yield-factor model of interest rates.
\newblock \emph{Mathematical Finance}, 6\penalty0 (4):\penalty0 379--406, 1996.

\bibitem[Duffie et~al.(2000)Duffie, Pan, and Singleton]{duffie2000transform}
D.~Duffie, J.~Pan, and K.~Singleton.
\newblock Transform analysis and asset pricing for affine jump-diffusions.
\newblock \emph{Econometrica}, 68\penalty0 (6):\penalty0 1343--1376, 2000.

\bibitem[Duffy(2006)]{duffy2013finite}
D.~J. Duffy.
\newblock \emph{Finite Difference Methods in Financial Engineering: A Partial Differential Equation Approach}.
\newblock Wiley Finance. John Wiley \& Sons, 2006.

\bibitem[Ekström and Vaicenavicius(2019)]{Ekström02012019}
E.~Ekström and J.~Vaicenavicius.
\newblock Monotonicity and robustness in wiener disorder detection.
\newblock \emph{Sequential Analysis}, 38\penalty0 (1):\penalty0 57--68, 2019.

\bibitem[Elliott and Siu(2009)]{Elliott2009}
R.~J. Elliott and T.~K. Siu.
\newblock On markov‐modulated exponential‐affine bond price formulae.
\newblock \emph{Applied Mathematical Finance}, 16\penalty0 (1):\penalty0 1--15, 2009.

\bibitem[Elliott et~al.(2005)Elliott, Chan, and Siu]{elliott2005option}
R.~J. Elliott, L.~Chan, and T.~K. Siu.
\newblock Option pricing and esscher transform under regime switching.
\newblock \emph{Annals of Finance}, 1\penalty0 (4):\penalty0 423--432, 2005.

\bibitem[Engel and Hamilton(1990)]{engel1990long}
C.~Engel and J.~D. Hamilton.
\newblock Long swings in the dollar: are they in the data and do markets know it.
\newblock \emph{American Economic Review}, 80\penalty0 (4):\penalty0 689--713, 1990.

\bibitem[Ernst and Peskir(2026)]{ernst2024gapeev}
P.~A. Ernst and G.~Peskir.
\newblock The gapeev--shiryaev conjecture.
\newblock \emph{The Annals of Applied Probability}, 36:\penalty0 3546--3565, 2026.

\bibitem[Gapeev and Shiryaev(2013)]{gapeev2013bayesian}
P.~V. Gapeev and A.~N. Shiryaev.
\newblock Bayesian quickest detection problems for some diffusion processes.
\newblock \emph{Advances in Applied Probability}, 45\penalty0 (1):\penalty0 164--185, 2013.

\bibitem[Gellert and Schl{\"o}gl(2021)]{gellert2021parameter}
K.~Gellert and E.~Schl{\"o}gl.
\newblock Parameter learning and change detection using a particle filter with accelerated adaptation.
\newblock \emph{Risks}, 9\penalty0 (12):\penalty0 228, 2021.

\bibitem[Glover and Peskir(2024)]{glover2024quickest}
K.~Glover and G.~Peskir.
\newblock Quickest detection problems for ornstein--uhlenbeck processes.
\newblock \emph{Mathematics of Operations Research}, 49\penalty0 (2):\penalty0 1045--1064, 2024.

\bibitem[Goldfeld and Quandt(1973)]{GOLDFELD19733}
S.~M. Goldfeld and R.~E. Quandt.
\newblock A markov model for switching regressions.
\newblock \emph{Journal of Econometrics}, 1\penalty0 (1):\penalty0 3--15, 1973.

\bibitem[Goutte et~al.(2017)Goutte, Ismail, and Pham]{goutte2017regime}
S.~Goutte, A.~Ismail, and H.~Pham.
\newblock Regime-switching stochastic volatility model: estimation and calibration to vix options.
\newblock \emph{Applied Mathematical Finance}, 24\penalty0 (1):\penalty0 38--75, 2017.

\bibitem[Gray(1996)]{gray1996modeling}
S.~F. Gray.
\newblock Modeling the conditional distribution of interest rates as a regime-switching process.
\newblock \emph{Journal of Financial Economics}, 42\penalty0 (1):\penalty0 27--62, 1996.

\bibitem[Hamilton(1989)]{hamilton1989new}
J.~D. Hamilton.
\newblock A new approach to the economic analysis of nonstationary time series and the business cycle.
\newblock \emph{Econometrica}, 57\penalty0 (2):\penalty0 357--384, 1989.

\bibitem[Heston(1993)]{heston1993closed}
S.~L. Heston.
\newblock A closed-form solution for options with stochastic volatility with applications to bond and currency options.
\newblock \emph{The Review of Financial Studies}, 6\penalty0 (2):\penalty0 327--343, 1993.

\bibitem[Jacod and Rosenbaum(2013)]{jacod2013quarticity}
J.~Jacod and M.~Rosenbaum.
\newblock Quarticity and other functionals of volatility: Efficient estimation.
\newblock \emph{The Annals of Statistics}, 41\penalty0 (3):\penalty0 1462--1484, 2013.

\bibitem[Johansen et~al.(2026)Johansen, Lange, and Skovmand]{johansen2026arithmetic}
M.~R. Johansen, N.~Lange, and D.~Skovmand.
\newblock Arithmetic models for oil futures, vanilla, and calendar spread options, 2026.
\newblock Accepted for publication in {\textit{Quantitative Finance}}.

\bibitem[Johnson and Peskir(2017)]{johnson2017quickest}
P.~Johnson and G.~Peskir.
\newblock Quickest detection problems for bessel processes.
\newblock \emph{The Annals of Applied Probability}, 27\penalty0 (2):\penalty0 1003--1056, 2017.

\bibitem[Kang(2014)]{kang2014estimation}
K.~H. Kang.
\newblock Estimation of state-space models with endogenous markov regime-switching parameters.
\newblock \emph{The Econometrics Journal}, 17\penalty0 (1):\penalty0 56--82, 2014.

\bibitem[Kokholm and Stisen(2015)]{kokholm2015joint}
T.~Kokholm and M.~Stisen.
\newblock Joint pricing of vix and spx options with stochastic volatility and jump models.
\newblock \emph{The Journal of Risk Finance}, 16\penalty0 (1):\penalty0 27--48, 2015.

\bibitem[Li(2013)]{LI201315}
J.~Li.
\newblock An unscented kalman smoother for volatility extraction: Evidence from stock prices and options.
\newblock \emph{Computational Statistics {\&} Data Analysis}, 58:\penalty0 15--26, 2013.

\bibitem[Ludkovski(2009)]{ludkovski2009simulation}
M.~Ludkovski.
\newblock A simulation approach to optimal stopping under partial information.
\newblock \emph{Stochastic Processes and their Applications}, 119\penalty0 (12):\penalty0 4061--4087, 2009.

\bibitem[Ludkovski(2012)]{ludkovski2012bayesian}
M.~Ludkovski.
\newblock Bayesian quickest detection in sensor arrays.
\newblock \emph{Sequential Analysis}, 31\penalty0 (4):\penalty0 481--504, 2012.

\bibitem[McGinnity and Irwin(2000)]{mcginnity2002multiple}
S.~McGinnity and G.~Irwin.
\newblock Multiple model bootstrap filter for maneuvering target tracking.
\newblock \emph{IEEE Transactions on Aerospace and Electronic Systems}, 36\penalty0 (3):\penalty0 1006--1012, 2000.

\bibitem[Mikosch and St{\u{a}}ric{\u{a}}(2004)]{mikosch2004changes}
T.~Mikosch and C.~St{\u{a}}ric{\u{a}}.
\newblock Changes of structure in financial time series and the garch model.
\newblock \emph{REVSTAT-Statistical Journal}, 2\penalty0 (1):\penalty0 41--73, 2004.

\bibitem[Naik(1993)]{naik1993option}
V.~Naik.
\newblock Option valuation and hedging strategies with jumps in the volatility of asset returns.
\newblock \emph{The Journal of Finance}, 48\penalty0 (5):\penalty0 1969--1984, 1993.

\bibitem[Nandi(1998)]{nandi1998important}
S.~Nandi.
\newblock How important is the correlation between returns and volatility in a stochastic volatility model? empirical evidence from pricing and hedging in the s\&p 500 index options market.
\newblock \emph{Journal of Banking \& Finance}, 22\penalty0 (5):\penalty0 589--610, 1998.

\bibitem[Pan(2002)]{pan2002jump}
J.~Pan.
\newblock The jump-risk premia implicit in options: Evidence from an integrated time-series study.
\newblock \emph{Journal of Financial Economics}, 63\penalty0 (1):\penalty0 3--50, 2002.

\bibitem[Papanicolaou and Sircar(2014)]{Papanicolaou01102014}
A.~Papanicolaou and R.~Sircar.
\newblock A regime-switching heston model for vix and s\&p 500 implied volatilities.
\newblock \emph{Quantitative Finance}, 14\penalty0 (10):\penalty0 1811--1827, 2014.

\bibitem[Peskir and Shiryaev(2006)]{peskir2006optimal}
G.~Peskir and A.~Shiryaev.
\newblock \emph{Optimal stopping and free-boundary problems}.
\newblock Springer, 2006.

\bibitem[Quandt(1958)]{quandt1958estimation}
R.~E. Quandt.
\newblock The estimation of the parameters of a linear regression system obeying two separate regimes.
\newblock \emph{Journal of the American Statistical Association}, 53\penalty0 (284):\penalty0 873--880, 1958.

\bibitem[Robert(2007)]{robert2007bayesian}
C.~P. Robert.
\newblock \emph{The Bayesian choice: from decision-theoretic foundations to computational implementation}.
\newblock Springer, 2007.

\bibitem[Shao(2018)]{shao2018invariant}
J.~Shao.
\newblock Invariant measures and euler--maruyama's approximations of state-dependent regime-switching diffusions.
\newblock \emph{SIAM Journal on Control and Optimization}, 56\penalty0 (5):\penalty0 3215--3238, 2018.

\bibitem[Shiryaev(1963)]{Shiryaev1963on}
A.~N. Shiryaev.
\newblock On optimum methods in quickest detection problems.
\newblock \emph{Theory of Probability \& Its Applications}, 8\penalty0 (1):\penalty0 22--46, 1963.

\bibitem[Shiryaev(2008)]{shiryaev2008optimal}
A.~N. Shiryaev.
\newblock \emph{Optimal stopping rules}.
\newblock Springer, 2008.

\bibitem[Siu and Elliott(2022)]{e69e7ccee7bf47feab824f343699e4bd}
T.~Siu and R.~Elliott.
\newblock American option pricing and filtering with a hidden regime-switching jump diffusion.
\newblock \emph{Journal of Derivatives}, 29\penalty0 (3):\penalty0 106--123, Mar. 2022.

\bibitem[Siu et~al.(2008)Siu, Yang, and Lau]{SIU2008295}
T.~K. Siu, H.~Yang, and J.~W. Lau.
\newblock Pricing currency options under two-factor markov-modulated stochastic volatility models.
\newblock \emph{Insurance: Mathematics and Economics}, 43\penalty0 (3):\penalty0 295--302, 2008.

\bibitem[Tong(1983)]{tong1983threshold}
H.~Tong.
\newblock \emph{Threshold models in non-linear time series analysis}.
\newblock Springer, 1983.

\bibitem[Trolle and Schwartz(2009)]{trolle2009unspanned}
A.~B. Trolle and E.~S. Schwartz.
\newblock Unspanned stochastic volatility and the pricing of commodity derivatives.
\newblock \emph{The Review of Financial Studies}, 22\penalty0 (11):\penalty0 4423--4461, 2009.

\bibitem[Tsaknaki et~al.(2025)Tsaknaki, Lillo, and Mazzarisi]{tsaknaki2025online}
I.-Y. Tsaknaki, F.~Lillo, and P.~Mazzarisi.
\newblock Online learning of order flow and market impact with bayesian change-point detection methods.
\newblock \emph{Quantitative Finance}, 25\penalty0 (2):\penalty0 307--322, 2025.

\end{thebibliography}

\appendix
\section{Oracle single change-point problem}\label{app:oracle}
In this section, we consider a slightly more general single change-point regime detection problem with a transient and an absorbing state. The observation process is $Y_t=(S_t,\nu_t)$, where
\begin{align*}
    {\rm d}S_t&=\mu S_t{\rm d}t+\sqrt{\nu_t}S_t{\rm d}W_t,\\
    {\rm d}\nu_t&=\alpha_{Z_t}(t,\nu_t){\rm d}t+\xi(t,\nu_t){\rm d}B_t,
    \qquad {\rm d}\langle B,W\rangle_t=\rho\,{\rm d}t.
\end{align*}
Here $\rho\in(-1,1)$, $\xi>0$, and the coefficients are locally Lipschitz on the state space $(0,\infty)^2$. We assume non-explosion and existence of a unique strong solution taking values in this state space, and that $(B,W)$ is independent of the regime process. In the Heston model, $\alpha_i(t,\nu)=\kappa(m_i-\nu)$ and $\xi(t,\nu)=\xi\sqrt{\nu}$. We write $\mathcal{F}^Y_t=\mathcal{F}^{S,\nu}_t$ for the observation filtration. Similar to adding time-dependent drift and diffusion, we will also allow for a time-inhomogeneous hidden Markov chain $(Z_t)$. Define the conditional change-point distribution $F(t)=\mathbb{P}(\theta\leq t\mid \theta>0)$, with density $f(t)$ and survival function $\overline{F}(t)$. That is the hazard rate of $(Z_t)$ entering the absorbing state is $\lambda(t)=f(t)/\overline{F}(t)$. The following is a generalization of the derivation in \cite{gapeev2013bayesian}.

Let $\pi=\mathbb{P}(\theta=0)$ and define the conditional measures $\mathbb{P}^s(\cdot)=\mathbb{P}(\cdot\mid \theta=s)$. The posterior probability process is given by 
\begin{equation}\label{eq:bayes_pi}
    \pi_t=\pi\frac{\mathrm{d}\mathbb{P}^0}{\mathrm{d}\mathbb{P}}\bigg\vert_{\mathcal{F}^{Y}_t}+(1-\pi)\int_0^t \frac{\mathrm{d}\mathbb{P}^s}{\mathrm{d}\mathbb{P}}\bigg\vert_{\mathcal{F}^{Y}_t}\mathrm{d}F(s),
\end{equation}
where $(\mathrm{d}\mathbb{P}^s/\mathrm{d}\mathbb{P})\vert \mathcal{F}^{Y}_t$ is the Radon-Nikodym derivative of $\mathbb{P}^s\vert\mathcal{F}^{Y}_t$ with respect to $\mathbb{P}\vert\mathcal{F}^{Y}_t$. Similarly,
\begin{equation}\label{eq:bayes_1-pi}
    1-\pi_t=(1-\pi)\int_t^\infty \frac{\mathrm{d}\mathbb{P}^s}{\mathrm{d}\mathbb{P}}\bigg\vert_{\mathcal{F}^{Y}_t}\mathrm{d}F(s)=(1-\pi)\overline{F}(t)\frac{\mathrm{d}\mathbb{P}^\infty}{\mathrm{d}\mathbb{P}}\bigg\vert_{\mathcal{F}^{Y}_t},
\end{equation}
since conditional on the filtration up to time $t$ then the measure $\mathbb{P}^s$ for $s>t$ is equal to the measure where $\theta$ never occurs. 

Next, we consider the posterior likelihood ratio process $(\phi_t)$ defined by $\phi_t=\pi_t/(1-\pi_t)$ which is 
\begin{equation*}
    \phi_t=\frac{1}{\overline{F}(t)} \left(\phi\frac{\mathrm{d}\mathbb{P}^0}{\mathrm{d}\mathbb{P}^\infty}\bigg\vert_{\mathcal{F}^{Y}_t}+\int_0^t\frac{\mathrm{d}\mathbb{P}^s}{\mathrm{d}\mathbb{P}^\infty}\bigg\vert_{\mathcal{F}^{Y}_t}\mathrm{d}F(s)\right),
\end{equation*}
where $\phi=\pi/(1-\pi)$. Thus, the posterior likelihood ratio process is given solely in terms of the Radon-Nikodym derivatives of the conditional measures.  Defining the likelihood process $L_t=(\mathrm{d}\mathbb{P}^0/\mathrm{d}\mathbb{P}^\infty)\vert \mathcal{F}^{Y}_t$ and assuming sufficient conditions for $L$ being a true martingale, 
see, e.g., \citet[][Theorem 1(c)]{cheridito2007market} for the Heston case, %
then Girsanov's theorem yields
\begin{align*}
    L_t &= \exp\left\{\int_0^t \eta_\rho(s,\nu_s){\rm d}B^{\infty,\perp}_s-\frac12\int_0^t\eta_\rho(s,\nu_s)^2{\rm d}s\right\}
\end{align*}
and $(\mathrm{d}\mathbb{P}^s/\mathrm{d}\mathbb{P}^\infty)\vert \mathcal{F}^{Y}_t=L_t/L_s$ for $s\leq t$, where $\eta_\rho$ is the correlation adjusted signal-to-noise ratio given by
\begin{equation*}
    \eta(t,\nu)=\frac{\alpha_2(t,\nu)-\alpha_1(t,\nu)}{\xi(t,\nu)},
    \qquad \eta_\rho(t,\nu)=\frac{\eta(t,\nu)}{\sqrt{1-\rho^2}},
\end{equation*}
and $(B^{\infty,\perp}_t)$ is a Brownian motion under $\mathbb{P}^\infty$, independent of $W$, given by
\begin{equation}\label{eq:bminf}
    {\rm d}B^{\infty,\perp}_t
    =\frac{1}{\xi(t,\nu_t)\sqrt{1-\rho^2}}
    \left[{\rm d}\nu_t-\alpha_1(t,\nu_t){\rm d}t
    -\frac{\rho\xi(t,\nu_t)}{\sqrt{\nu_t}}
    \left(\frac{{\rm d}S_t}{S_t}-\mu\,{\rm d}t\right)\right]
\end{equation}
Thus, the posterior likelihood ratio dynamics follow from Ito's formula,
\begin{equation*}
    {\rm d}\phi_t=\lambda(t)(1+\phi_t){\rm d}t+\eta_\rho(t,\nu_t)\phi_t {\rm d}B^{\infty,\perp}_t,
\end{equation*}
and the posterior probability process satisfies
\begin{equation*}
    {\rm d}\pi_t=\lambda(t)(1-\pi_t){\rm d}t+\eta_\rho(t,\nu_t)\pi_t(1-\pi_t){\rm d}I_t,
\end{equation*}
where the innovation process is
\begin{equation*}
    {\rm d}I_t=
    \frac{1}{\xi(t,\nu_t)\sqrt{1-\rho^2}}
    \left[{\rm d}\nu_t-\bar\alpha(t,\nu_t,\pi_t){\rm d}t
    -\frac{\rho\xi(t,\nu_t)}{\sqrt{\nu_t}}
    \left(\frac{{\rm d}S_t}{S_t}-\mu\,{\rm d}t\right)\right],
\end{equation*}
where
\begin{equation*}
    \bar\alpha(t,\nu,\pi)=\alpha_1(t,\nu)+\pi\big(\alpha_2(t,\nu)-\alpha_1(t,\nu)\big).
\end{equation*}
The pair $(W,I)$ is a two-dimensional uncorrelated $(\mathcal{F}^{Y}_t)$-Brownian motion under $\mathbb{P}$.
Writing the variance process in terms of the innovation process, 
\begin{equation*}
    {\rm d}\nu_t=\bar\alpha(t,\nu_t,\pi_t){\rm d}t
    +\xi(t,\nu_t)\big(\rho\,{\rm d}W_t+\sqrt{1-\rho^2}\,{\rm d}I_t\big),
\end{equation*}
we see that $(t,\nu_t,\pi_t)$ is a Markov process, since its coefficients and the decision costs do not depend on the stock level $S_t$. Thus, we can consider the Markovian version of \eqref{eq:OSP_general} by conditioning on the starting point
\begin{equation*}
    V(t,\nu,\pi)=\inf_{\tau\leq T-t}\mathbb{E}\left[g(\pi_{t+\tau})+\int_0^\tau h(\pi_{t+u})\,{\rm d}u\;\Big\vert\; \nu_t=\nu,\pi_t=\pi\right].
\end{equation*}
It follows from standard arguments in general optimal stopping theory \citep[][Corollary 2.9]{peskir2006optimal} that the state space can be partitioned into the so-called continuation set $C$ and stopping set $D$ given by 
\begin{equation*}
    C = \{(t,\nu,\pi)\mid V(t,\nu,\pi)<g(\pi)\},\quad\text{and}\quad D = \{(t,\nu,\pi)\mid V(t,\nu,\pi)=g(\pi)\},
\end{equation*}
and that the optimal stopping time is given by the first entry time to the set $D$, i.e., 
\begin{equation*}
    \tau^*=\inf\{s\in[0,T-t]\mid(t+s,\nu_{t+s},\pi_{t+s})\in D\}\wedge(T-t)
\end{equation*}
provided that $\mathbb{P}(\tau^*<\infty\mid \nu_t=\nu,\pi_t=\pi)=1$ for all $t,\nu,$ and $\pi$ in the state space. 
Some properties of the stopping set $D$ are easily inferred from considering the Lagrange formulation. At time zero, the quickest detection value can be written under the $\mathbb{P}^\infty$ measure as
\begin{equation}\label{eq:bayes_phi}
    V = (1-\pi)\left(b+\inf_{\tau\leq T}\mathbb{E}^\infty\left[\int_0^\tau \big(c\overline{F}(t)\phi_t -bf(t)\big)\;\mathrm{d}t\right]\right).
\end{equation}
The infimum inside the parentheses is nondecreasing in the initial position of $(\phi_t)$, and furthermore, before the terminal time it is never optimal to stop when 
\begin{equation*}
    c\overline{F}(t)\phi_t -bf(t)<0.
\end{equation*}
Defining $\underline{\pi}(t)=b\lambda(t)/(c+b\lambda(t))$, the latter implies that 
\begin{equation*}
    \{(t,\nu,\pi)\mid t<T,\ \pi<\underline{\pi}(t)\}\subseteq C,
\end{equation*}
while the former implies that if $(t,\nu,\pi)\in D$ then $(t,\nu,\pi+\varepsilon)\in D$ for any $\varepsilon\in[0,1-\pi]$. This means that there exists a function $\beta:[0,T]\times(0,\infty)\to [0,1]$ such that
\begin{equation*}
    D = \{(t,\nu,\pi)\mid \pi\geq \beta(t,\nu)\},
\end{equation*}
where $\beta(t,\nu)\geq \underline{\pi}(t)$ for all $t<T$ and $\nu\in(0,\infty)$.

In order to characterize the optimal stopping boundary further, one has to establish the smooth fit property of the value function.
We restrict the following discussion to the quickest detection problem in the time-homogeneous Heston single-change-point problem on an infinite time horizon $T=\infty$ where the change-point only affects the mean-reversion level and $m_1<m_2$. Since the problem is time-homogeneous, we drop the $t$ dependence.

First, recall that $(\nu_t,\pi_t)$ is a strong Feller process. Indeed, for $\rho\neq0$, the covariance matrix is strictly positive on $(0,\infty)\times(0,1)$. Thus, the diffusion is locally uniformly elliptic and $(\nu_t,\pi_t)$ is a strong Feller process. For $\rho=0$, the process satisfies the no trap conditions of \cite{ernst2024gapeev} with analytic coefficients, and so $(\nu_t,\pi_t)$ is a strong Feller process. The dynamic programming principle gives $\mathcal{L}V=-h$ in the continuation set. Standard local elliptic regularity for $\rho\neq0$, and the hypoelliptic argument in \citet[][proof of Theorem~6]{ernst2024gapeev} for $\rho=0$, imply that $V$ is twice continuously differentiable in $C$.

Next, we show that $\nu\mapsto \beta(\nu)$ is non-increasing. To this end, it is enough to show that $\nu\mapsto V(\nu,\pi)$ is non-decreasing for any fixed $\pi\in(0,1)$. We adapt the methodology of \cite{Ekström02012019} to our model. Let $v_1<v_2$ and denote by $(S^i_t,\nu^i_t)$ the observation path when $S^i_0=s$ and $\nu^i_0=v_i$ for $i=1,2$. Take a correlated 2-dimensional Brownian motion $(B'_t,W'_t)$ with $\langle B',W'\rangle_t=\rho t$ independent of $(B_t,W_t,Z_t)$. Let $(U_t)$ be the non-negative solution to
\begin{equation*}
    dU_t=-\kappa U_tdt+\xi\sqrt{U_t}dB'_t,\qquad U_0=v_2-v_1,
\end{equation*}
with absorption at $0$. Now define 
\begin{equation*}
    \tilde{\nu}^2_t=\nu^1_t+U_t,\qquad \tilde{S}^2_t=S^1_t \exp\left(\int_0^t \sqrt{U_s}dW'_s-\frac{1}{2}\int_0^t U_s ds\right).
\end{equation*}
By an application of Ito, it is not hard to see that $(\theta,(\tilde{S}^2_t),(\tilde{\nu}^2_t))\overset{d}{=}(\theta,(S^2_t),(\nu^2_t))$. Thus, if we write $J(\tau)=\mathbb{E}\left[b1_{\{\tau<\theta\}}+c(\tau-\theta)^+\right]$, we have that 
\begin{equation*}
    V(v_1,\pi)=\inf_{\tau\in\mathcal{T}^{S^1,\nu^1}}J(\tau)=\inf_{\tau\in\mathcal{T}^{S^1,\nu^1,B',W'}}J(\tau)\leq \inf_{\tau\in\mathcal{T}^{\tilde{S}^2,\tilde{\nu}^2}}J(\tau)=V(v_2,\pi),
\end{equation*}
where the second equality holds because independent randomization cannot reduce the minimum expected loss and $\mathcal{T}^{\tilde{Y}}$ is the set of admissible stopping times with respect to the augmented filtration generated by $(\tilde{Y}_t)$. This could also be realized in the $\rho=0$ case by applying Theorem 6 in \cite{ernst2024gapeev}.

Next, we need to show the probabilistic regularity of the boundary. Let $\nu_0=\nu$ and $\pi_0=\pi$. Let $(\nu,\pi)\in\partial C\cap((0,\infty)\times(0,1))$. Since $D$ is a closed set, then $(\nu,\pi)\in D$. Since $D$ is given by a non-increasing function $\beta$, then $(\nu,\infty)\times(\pi,1)\subseteq D^\circ$. Defining
\begin{equation*}
    \sigma_{D^\circ}=\inf\{t>0\mid (\nu_t,\pi_t)\in D^\circ\},
\end{equation*}
then the preceding analysis gives 
\begin{equation*}
    \mathbb P_{(\nu,\pi)}(\sigma_{D^\circ}=0)\geq \limsup_{t\searrow0}\mathbb P_{(\nu,\pi)}(\nu_t>\nu,\pi_t>\pi).
\end{equation*}
Note that in the small-time limit, we have 
\begin{equation*}
    \left(\frac{\nu_t-\nu}{\xi\sqrt{\nu}\sqrt{t}},\frac{\pi_t-\pi}{\eta_\rho(\nu)\pi(1-\pi)\sqrt{t}}\right)\overset{d}{\to}\left(\rho N_1+rN_2,N_2\right),
\end{equation*}
where $N_1$ and $N_2$ are independent standard Gaussian random variables and $r=\sqrt{1-\rho^2}$. Thus, 
\begin{align*}
    \lim_{t\searrow0}\mathbb P_{(\nu,\pi)}(\nu_t>\nu,\pi_t>\pi)&=\mathbb P(\rho N_1+rN_2>0,N_2>0)=\frac{1}{4}+\frac{1}{2\pi}\arcsin(r)>0,
\end{align*}
and so $\mathbb P_{(\nu,\pi)}(\sigma_{D^\circ}=0)>0$. An application of the Blumenthal 0--1 law yields that the $\partial C$ is probabilistically regular for the stopping set $D$.

Applying the smooth-fit criterion of \citet[][Theorem 8]{de2020global}, using the strong Feller property and probabilistic regularity of the boundary, yields continuous differentiability of \(V\) across the interior free boundary.
Thus, the value satisfies the free-boundary problem
\begin{align*}
    \mathcal{L}V(\nu,\pi)&=-h(\pi) && (\nu,\pi)\in C \\
    V(\nu,\pi)&=g(\pi) && (\nu,\pi)\in D \\
    V_\pi(\nu,\pi)&=g'(\pi) && (\nu,\pi)\in\partial D\\
    V_{\nu}(\nu,\pi)&=0 && (\nu,\pi)\in\partial D
\end{align*}
where $V_\pi$ and $V_{\nu}$ are the first partial derivatives of $V$, and $\mathcal{L}$ is the infinitesimal generator of $(\nu_t,\pi_t)$, that is 
\begin{equation*}
    \begin{aligned}
    \mathcal{L}\varphi(\nu,\pi) =& \big(\alpha_1(\nu)+\pi\big(\alpha_2(\nu)-\alpha_1(\nu)\big)\big)\varphi_{\nu}(\nu,\pi)+\lambda(1-\pi)\varphi_\pi(\nu,\pi)\\ &+\xi(\nu)\eta(\nu)\pi(1-\pi)\varphi_{\nu\pi}(\nu,\pi)
    +\frac{1}{2}\xi^2(\nu) \varphi_{\nu\nu}(\nu,\pi)+\frac{1}{2}\big(\eta_\rho(\nu)\pi(1-\pi)\big)^2\varphi_{\pi\pi}(\nu,\pi).
    \end{aligned}
\end{equation*}
For methods of establishing continuity and uniqueness of monotone optimal stopping boundaries using smooth fit, see \cite{peskir2006optimal}.

\section{Kushner-Stratonovich filters}\label{app:filters}
We derive the Kushner-Stratonovich filter and show that the measure-valued optimal stopping problem can be handled by the methodology in \cite{chiarolla2016optimal}. 
We consider the problem under operational time similar to \cite{AIHARA20031681}. Consider the stochastic clock $\tilde{A}_t=\int_0^t\nu_u{\rm d}u$ which is strictly increasing by the Feller condition assumed and so it has an inverse which we denote $A_s$. Defining $\tilde{Y}_s=\log S_{A_s}$, $\tilde{Z}_s =Z_{A_s}$, and $\tilde{X}_s = \nu_{A_s}$, we have 
\begin{align}
    {\rm d}\tilde{Y}_s &= \left(\frac{\mu}{\tilde{X}_s}-\frac{1}{2}\right){\rm d}s+{\rm d}\tilde{W}_s,\\
    {\rm d}\tilde{X}_s&=\kappa\left(\frac{m_{\tilde{Z}_s}}{\tilde{X}_s}-1\right){\rm d}s+\xi {\rm d}\tilde{B}_s
\end{align}
where $\tilde{W}_s$ and $\tilde{B}_s$ are correlated Brownian motions with $\mathrm{d}\langle \tilde{W},\tilde{B}\rangle_s=\rho\,{\rm d}s$, and we assume the $(\tilde{X}_0,\tilde{Z}_0)$ prior has an $L^2((0,\infty)\times\{1,2\})$ density. Furthermore, the hazard rate becomes dependent on the latent variable $\tilde{\lambda}(\nu)=\lambda/\nu$. It should be clear that $\hat{\nu}_s=\mathbb{E}[\tilde{X}_s\mid\mathcal{F}^{\tilde{Y}}_s]$ is not Markov, and more importantly $(\hat{X}_s,\pi_{A_s})$ is not Markov with respect to its own filtration. 

If we instead consider the full posterior distribution of the latent factor and the hidden regime, then we indeed recover the strong Markov property. Namely, let $\tilde\Pi_s$ denote the conditional probability measure of $(\tilde X_s,\tilde Z_s)$ given $\mathcal F_s^{\tilde Y}$ and write its action on a test function as
\begin{equation*}
    \tilde{\Pi}_s(\varphi)=\mathbb{E}[\varphi(\tilde{X}_s,\tilde{Z}_s)\mid\mathcal{F}^{\tilde{Y}}_s]
\end{equation*}
for measurable test-functions $\varphi$. The filter $\tilde{\Pi}$ is strong Markov, which we can show by standard stochastic filtering theory \citep[see, e.g.,][]{bain2009fundamentals}.
The infinitesimal generator of $(\tilde{X}_s,\tilde{Z}_s)$ is given by 
\begin{equation*}
    \tilde{\mathcal{L}}\varphi(\nu,z)= \kappa\left(\frac{m_z}{\nu}-1\right)\varphi_\nu(\nu,z)+\frac{1}{2}\xi^2\varphi_{\nu\nu}(\nu,z)+(2-z)\frac{\lambda}{\nu}\big(\varphi(\nu,2)-\varphi(\nu,1)\big).
\end{equation*}
Then the filter 
is given by the unnormalized conditional distribution $(\rho_t(\cdot))$ through
\begin{equation*}
    \tilde{\Pi}_s(\varphi) = \frac{\rho_s(\varphi)}{\rho_s(1)}.
\end{equation*}
The $(\rho_s(\cdot))$ process satisfies the Zakai equation
\begin{equation*}
    {\rm d}\rho_s(\varphi)=\rho_s(\tilde{\mathcal{L}}\varphi){\rm d}s+\rho_s\Big(\varphi f+\rho\xi\varphi_v\Big){\rm d}\tilde{Y}_s,
\end{equation*}
and $(\tilde{I}_s)$ is the innovation process %
\begin{equation*}
    {\rm d}\tilde{I}_s={\rm d}\tilde{Y}_s-\tilde{\Pi}_s(f){\rm d}s,
\end{equation*}
which is a Brownian motion under $(\mathcal{F}^{\tilde{Y}}_s)$.
This in turn yields the Kushner-Stratonovich equation
\begin{equation*}
        {\rm d}\tilde{\Pi}_s(\varphi)=\tilde{\Pi}_s(\tilde{\mathcal{L}}\varphi){\rm d}s+\Big(\tilde{\Pi}_s(\varphi f)-\tilde{\Pi}_s(\varphi)\tilde{\Pi}_s(f)+\rho\xi\,\tilde{\Pi}_s(\varphi_v)\Big){\rm d}\tilde{I}_s,
\end{equation*}
for the filter.
Reverting the time change, we arrive at 
\begin{equation*}
        {\rm d}\Pi_t(\varphi)=\Pi_t(\mathcal{L}\varphi){\rm d}t+\Pi_t(\sqrt{\nu})\Big(\Pi_t(\varphi f)-\Pi_t(\varphi)\Pi_t(f)+\rho\xi\,\Pi_t(\varphi_v)\Big){\rm d}I_t,
\end{equation*}
where 
$(I_t)$ is a Brownian motion under $(\mathcal{F}^{Y}_t)$ given by $\sqrt{\nu_t}{\rm d}I_t={\rm d}\tilde{I}_{\tilde{A}_t}$.

We finish this appendix by showing that the filter $(\Pi_t)$ is itself a strong Markov process in the Heston case. Let $\tau$ be an $(\mathcal F_t^S)$-stopping time and define the normalized future observation $X^{\tau}_s=S_{\tau+s}/S_\tau$, $s\geq0$. Since the dynamics of $(X^{\tau},\nu_{\tau+\cdot}, Z_{\tau+\cdot})$ do not depend on $S_\tau$, the strong Markov property of $(S,\nu,Z)$ implies that, conditional on $\mathcal F_\tau^{S,\nu,Z}$, their future law depends on the past only through $(\nu_\tau,Z_\tau)$. Moreover, as in \eqref{eq:filter_markov}, filtering can be restarted from the current conditional distribution $\Pi_\tau$. Hence, for every bounded measurable $f:\mathcal P(E)\to\mathbb R$ and $s\geq0$, there exists a measurable function $K_s$ such that $\mathbb E[f(\Pi_{\tau+s})\mid\mathcal F_\tau^{S,\nu,Z}]
=K_s(\nu_\tau,Z_\tau,\Pi_\tau)$ and so by the tower property, we have
\[
\mathbb E\left[
    f(\Pi_{\tau+s})
    \mid\mathcal F_\tau^S
\right]=\mathbb E\left[\mathbb E\left[
    f(\Pi_{\tau+s})
    \mid\mathcal F_\tau^{S,\nu,Z}
\right]\mid \mathcal{F}_\tau^S\right]
=
\int_E K_s(\nu,z,\Pi_\tau)\,
       \Pi_\tau({\rm d}\nu,{\rm d}z).
\]
Thus, the conditional law of $\Pi_{\tau+s}$ given $\mathcal F_\tau^S$ depends on the past only through $\Pi_\tau$, and therefore $(\Pi_t)$ is strong Markov.

\section{Bootstrap particle filter}\label{app:pf}
We describe a simple particle filter for regime switching models like \cite{mcginnity2002multiple} with the usual effective sample size resampling rule. The measurement equation in \eqref{eq:obs_rec} induces an observation likelihood for $\bm{Y}_{t+\Delta}$ given propagated latent factors and hidden regimes. We write 
\begin{equation*}
    \bm{Y}_{t+\Delta}\sim q_{\Delta}(\cdot\mid \bm{Y}_{t},\bm{X}_{t+\Delta},Z_{t+\Delta}),
\end{equation*}
where, for instance, %
in an Euler-Maruyama scheme, $q_{\Delta}(\cdot\mid \bm{Y}_{t},\bm{X}_{t+\Delta},Z_{t+\Delta})$ is Gaussian. 
We approximate $\Pi_t$ by the empirical measure $\widehat{\Pi}_t$ on $M$ particles $(\bm{X}_t^{(m)},Z_t^{(m)})_{m=1}^M$ with weights $(w_t^{(m)})_{m=1}^M$, that is, for every test function $\varphi$,
\begin{equation*}
    \widehat{\Pi}_t(\varphi)=\sum_{m=1}^M w_t^{(m)}\varphi(\bm X_t^{(m)},Z_t^{(m)})\approx\Pi_t(\varphi),
\end{equation*}
where $\sum_{m=1}^Mw^{(m)}_t=1$ and $w^{(m)}_t\geq0$. The particles can be initiated in the prior of $(\bm{X}_0,Z_0)$ with equal weights. A particle filter step from $t$ to $t+\Delta$ where a new $\bm{Y}_{t+\Delta}$ is observed consists of the following steps:
\begin{enumerate}
    \item \textbf{Propagation}: Sample $(\bm{X}^{(m)}_{t+\Delta},Z^{(m)}_{t+\Delta})\vert (Y_t,\bm{X}^{(m)}_{t},Z^{(m)}_{t})$ by \eqref{eq:lat_rec}--\eqref{eq:hid_rec}.
    \item \textbf{Weight update}: update unnormalized weights via 
    \begin{equation*}
        \widetilde{w}^{(m)}_{t+\Delta}=w^{(m)}_{t}q_{\Delta}(\bm{Y}_{t+\Delta}\mid \bm{Y}_{t},\bm{X}_{t+\Delta}^{(m)},Z_{t+\Delta}^{(m)}),
    \end{equation*}
    and normalize $w^{(m)}_{t+\Delta}=\widetilde{w}^{(m)}_{t+\Delta}/\sum_{l=1}^M\widetilde{w}^{(l)}_{t+\Delta}$.
    \item \textbf{Resampling}: Calculate effective sample size $(\mathrm{ESS})$
    \begin{equation*}
        \mathrm{ESS}_{t+\Delta} = \frac{1}{\sum_{m=1}^M (w_{t+\Delta}^{(m)})^2}.
    \end{equation*}
    If $\mathrm{ESS}_{t+\Delta} < a M$ for a chosen threshold $a\in(0,1)$, resample particles according to their weights and reset weights to $1/M$.
\end{enumerate}
The last step combats particle impoverishment by discarding low weight particles if the effective sample size is sufficiently low.

For the regression in PFMCDP, we gather the regression features before the resampling step.
Define the raw moments
\begin{equation*}
\widehat\pi_t^{(j)}=\sum_{m=1}^M w_t^{(m)}1_{\{Z_t^{(m)}=j\}},\qquad
\widehat m_t=\sum_{m=1}^M w_t^{(m)}X_t^{(m)},\qquad
\widehat q_t=\sum_{m=1}^M w_t^{(m)}(X_t^{(m)})^2,\qquad
\widehat v_t=\widehat q_t-\widehat m_t^2,
\end{equation*}
and, for each regime $j$, we use the regime dependent moments
\begin{equation*}
\widehat m_t^{(j)}=\sum_{m=1}^M w_t^{(m)}X_t^{(m)}1_{\{Z_t^{(m)}=j\}},\qquad
\widehat q_t^{(j)}=\sum_{m=1}^M w_t^{(m)}(X_t^{(m)})^2 1_{\{Z_t^{(m)}=j\}},\qquad
\widehat v_t^{(j)}=\widehat q_t^{(j)}-(\widehat m_t^{(j)})^2,
\end{equation*}
since the regime conditional moments can become numerically unstable when few particles are in a given regime. The baseline covariates used in the Vasicek example are 
\begin{equation*}
    \left(Y_t,(\widehat \pi^{(j)}_t)_{j\in\mathcal{S}}, \widehat m_t, \widehat v_t, ( \widehat m^{(j)}_t,\widehat q^{(j)}_t)_{j\in\mathcal{S}_2}\right),
\end{equation*}
and the non-linear terms $\widehat m_t^2$, $((\widehat\pi^{(j)}_t)^2)_{j\in\mathcal{S}}$, and $(\widehat m\widehat\pi^{(j)}_t)_{j\in\mathcal{S}_2}$. Note that $(\widehat \pi^{(j)}_t)_{j\in\mathcal{S}}$ sums to one, so we get the intercept term.  
For the Heston example, we use the same covariates, except we drop the observation process since the filter is Markovian by itself, and we replace $\widehat \pi^{(2)}_t$ with $\widehat \pi^{\mathrm{CL}}_t$ given in \eqref{eq:heston_cl_posterior} and $\widehat \pi^{(1)}_t$ with $1-\widehat \pi^{\mathrm{CL}}_t$, while retaining the direct particle estimate $\widehat \pi^{(1)}_t$ as one additional covariate. %
The particle ancestry needed to calculate $\widehat \pi^{\mathrm{CL}}_t$ is then also resampled in the resampling step described above.

\section{Vasicek with regime switching market price of risk}\label{app:vasicek}
Under the risk-neutral measure $\mathbb{Q}$ the latent short rate has regime switching mean level
\begin{equation*}
    {\rm d}X_t=\kappa(m^\mathbb{Q}_{Z_t}-X_t)dt+\sigma{\rm d}B^\mathbb{Q}_t, %
\end{equation*}
where 
\begin{equation*}
    m^\mathbb{Q}_{Z_t} = \theta-\sigma\frac{\zeta_{Z_t}}{\kappa}.%
\end{equation*}
The price of a zero-coupon bond is 
\begin{equation*}
    p_i(t,x)=\mathbb{E}^\mathbb{Q}\left[e^{-\int_t^T X_s {\rm d}s}\big \vert X_t=x, Z_t=i\right],
\end{equation*}
with $p_i(T,x)=1$. The price $p$ solves the PDE
\begin{equation*}
    \partial_t p_i+\kappa(m^{\mathbb{Q}}_i-x)\partial_xp_i+\frac{1}{2}\sigma^2\partial_{xx}p_i+\sum_{j\in\mathcal{S}}q_{ij}p_j-xp_i=0.%
\end{equation*}
An ansatz of $p_i(t,x)=A_i(T-t)e^{-B(T-t)x}$ yields
\begin{equation*}
    B(\tau)=\frac{1-e^{-\kappa\tau}}{\kappa}
\end{equation*}
and $\bm{A}(\cdot)=(A_i(\cdot))_{i\in\mathcal{S}}$ solves the $n$--dimensional ODE
\begin{equation*}
    \nabla \bm{A}(\tau)=\bigg(\mathbf{Q}-\mathrm{diag}\Big(\kappa m^\mathbb{Q}_i B(\tau)-\frac{1}{2}\sigma^2B^2(\tau)\Big)\bigg)\bm{A}(\tau), %
\end{equation*}
which can be solved by standard numerical techniques.
\end{document}